\documentclass[12pt, letterpaper]{JHEP3}

\usepackage{amssymb, amsmath, amsopn, amsthm}

\usepackage{epsfig}

\newcommand{\eref}[1]{(\ref{#1})}

\newcommand{\fref}[1]{Figure~\ref{#1}}
\newcommand{\cref}[1]{Chapter~\ref{#1}}
\newcommand{\beq}{\begin{equation}}
\newcommand{\eeq}{\end{equation}}
\newcommand{\ba}{\begin{array}}
\newcommand{\ea}{\end{array}}
\newcommand{\bcenter}{\begin{center}}
\newcommand{\ecenter}{\end{center}}

\def\IB{\relax\hbox{$\inbar\kern-.3em{\rm B}$}}
\def\IC{\relax\hbox{$\inbar\kern-.3em{\rm C}$}}
\def\ID{\relax\hbox{$\inbar\kern-.3em{\rm D}$}}
\def\IE{\relax\hbox{$\inbar\kern-.3em{\rm E}$}}
\def\IF{\relax\hbox{$\inbar\kern-.3em{\rm F}$}}
\def\IG{\relax\hbox{$\inbar\kern-.3em{\rm G}$}}
\def\IGa{\relax\hbox{${\rm I}\kern-.18em\Gamma$}}
\def\IH{\relax{\rm I\kern-.18em H}}
\def\IK{\relax{\rm I\kern-.18em K}}
\def\IL{\relax{\rm I\kern-.18em L}}
\def\IP{\relax{\rm I\kern-.18em P}}
\def\IR{\relax{\rm I\kern-.18em R}}
\def\IZ{\relax\ifmmode\mathchoice
{\hbox{\cmss Z\kern-.4em Z}}{\hbox{\cmss Z\kern-.4em Z}}
{\lower.9pt\hbox{\cmsss Z\kern-.4em Z}}
{\lower1.2pt\hbox{\cmsss Z\kern-.4em Z}}\else{\cmss Z\kern-.4em Z}\fi}
\def\II{\relax{\rm I\kern-.18em I}}

\def\sCC{{\kern 0.27em\vrule height1.45ex width0.03em depth0em
          \kern-0.30em\rm C}}
\def\C{{\mathchoice
  {\sCC}
  {\sCC}
  {\kern 0.225em \vrule height1.05ex width0.025em depth0em \kern-0.25em \rm C}
  {\kern 0.180em \vrule height0.78ex width0.02em depth0em \kern-0.2em \rm C}
        }}
\def\sHH{{\rm I\kern-.16em{}H}}
\def\H{{\mathchoice
  {\sHH}
  {\sHH}
  {\rm I\kern-.13em{}H}
  {\rm I\kern-.13em{}H} }}
\def\sNN{{\rm I\kern-.16em{}N}}
\def\N{{\mathchoice
  {\sNN}
  {\sNN}
  {\rm I\kern-.12em{}N}
  {\rm I\kern-.10em{}N} }}
\def\sPP{{\rm I\kern-.16em{}P}}
\def\P{{\mathchoice
  {\sPP}
  {\sPP}
  {\rm I\kern-.12em{}P}
  {\rm I\kern-.10em{}P} }}
\def\sQQ{{\kern 0.27em \vrule height1.45ex width0.03em depth0em
          \kern-0.30em \rm Q}}
\def\Q{{\mathchoice
        {\sQQ}
        {\sQQ}
  {\kern 0.225em \vrule height1.05ex width0.025em depth0em \kern-0.25em \rm Q}
  {\kern 0.180em \vrule height0.78ex width0.020em depth0em \kern-0.20em \rm Q}
        }}
\def\sRR{{\rm I\kern-0.16em{}R}}
\def\R{{\mathchoice
  {\sRR}
  {\sRR}
  {\rm I\kern-0.12em{}R}
  {\rm I\kern-0.10em{}R} }}
\def\sZZ{{\rm Z\kern-0.32em{}Z}}
\def\Z{{\mathchoice
  {\sZZ}
  {\sZZ} 
  {\rm Z\kern-0.3em{}Z}     
  {\rm Z\kern-0.25em{}Z} }}  
\def\ZZZ{{\rm Z\kern-0.24em{}Z}}
\def\sII{{\rm I\kern-0.16em{}I}}
\def\I{{\mathchoice
  {\sII}
  {\sII}
  {\rm I\kern-0.12em{}I}
  {\rm I\kern-0.10em{}I} }}

\def\Tr{{\rm Tr}}

\def\inbar{\,\vrule height1.5ex width.4pt depth0pt}
\font\cmss=cmss10 \font\cmsss=cmss10 at 7pt

\def\smiley{\hbox{\large$\bigcirc$\hspace{-0.80em}\raise.2ex
\hbox{$\cdot\cdot$}\kern-.61em\lower.2ex\hbox{\scriptsize$\smile$}}\ }
\def\frowny{\hbox{\large$\bigcirc$\hspace{-0.80em}\raise.2ex
\hbox{$\cdot\cdot$}\kern-.635em\lower.2ex\hbox{\scriptsize$\frown$}}\ }

\def\I{{\rlap{1} \hskip 1.6pt \hbox{1}}}

\makeatletter
\let\hangafter\@hangfrom
\makeatother

\def\makeatletter{\catcode`\@=11}
\makeatletter
\def\mathbox#1{\hbox{$\m@th#1$}}%
\def\math@ccstyles#1#2#3#4#5#6#7{{\leavevmode
     \setbox0\mathbox{#6#7}%
     \setbox2\mathbox{#4#5}%
     \dimen@ #3%
     \baselineskip\z@\lineskiplimit#1\lineskip\z@
     \vbox{\ialign{##\crcr
            \hfil \kern #2\box2 \hfil\crcr
            \noalign{\kern\dimen@}%
            \hfil\box0\hfil\crcr}}}}
\def\mathaccstyles{\math@ccstyles\maxdimen}
\def\maththroughstyles{\math@ccstyles{-\maxdimen}}
\def\unity%
{\maththroughstyles{.45\ht0}\z@\displaystyle {\mathchar"006C}\displaystyle 1}

\title{Free Energy vs Sasaki-Einstein Volume for Infinite Families of M2-Brane Theories}

\author{Antonio Amariti$^1$ and Sebasti\'an Franco$^{2,3}$

\\

\vspace{0.05cm}
~\\
$^1$ Department of Physics, University of California \\
San Diego, La Jolla, CA 92093-0354, USA \\
\vspace{0.07cm}

$^2$ Theory Group, SLAC National Accelerator Laboratory \\
Menlo Park, CA 94309, USA \\
\vspace{0.07cm}

$^3$ Institute for Particle Physics Phenomenology, Department of Physics \\
Durham University, Durham DH1 3LE, United Kingdom \\

\vspace{0.05cm}

\email{amariti@physics.ucsd.edu, sfranco@slac.stanford.edu}\\

}

\abstract{We investigate infinite families of 3d $\mathcal{N}=2$ superconformal Chern-Simons quivers with an arbitrarily large number of gauge groups arising on M2-branes over toric CY$_4$'s. These theories have the same matter content and superpotential of those on D3-branes probing cones over $L^{a,b,a}$ Sasaki-Einstein manifolds. For all these infinite families, we explicitly show the correspondence between the free energy $F$ on $S^3$ and the volume of the 7-dimensional base of the associated CY$_4$, even before extremization. Our results add to those existing in the literature, providing further support for the correspondence. 
We develop a lifting algorithm, based on the Type IIB realization of these theories, that takes from CY$_3$'s to CY$_4$'s and we use it to efficiently generate the models studied in the paper. We also introduce a procedure, based on the mapping between extremal points in the toric diagram (GLSM fields) 
and chiral fields in the quiver, which systematically translates symmetries of the toric diagram into constraints of the trial R-charges of the quiver, beyond those arising from marginality of the superpotential. This method can be exploited for reducing the dimension of the space of trial R-charges over which the free energy is maximized. Finally, we show that in all the infinite families we consider $F^2$ can be expressed, even off-shell, as a quartic function in R-charges associated to certain 5-cycles. This suggests that a quartic formula on R-charges, analogous to a similar cubic function for the central charge $a$ in 4d, exists for all toric toric CY$_4$'s and we present some ideas regarding its general form.
}

\preprint{UCSD-PTH-12-06, SLAC-PUB-14968 \\ IPPP/12/23, DCPT/12/46}

\def\be{\begin{equation}}

\def\ee{\end{equation}}

\def\bea{\begin{eqnarray}}

\def\eea{\end{eqnarray}}

\begin{document}

\tableofcontents

\section{Introduction}

\label{section_introduction}

In recent years we have witnessed remarkable progress in the study of 3d superconformal field theories (SCFTs) on two tightly interconnected fronts. Progress in any of the two directions has fueled new advances in the other one.

The first front involves the determination of SCFTs describing the low energy dynamics of M2-branes. Following the seminal ideas of \cite{Bagger:2006sk,Gustavsson:2007vu,Bagger:2007jr,Bagger:2007vi}, which culminated with the construction of a 3d superconformal Chern-Simons (CS) theory with maximal $\mathcal{N}=8$ supersymmetry (SUSY), a theory describing $N$ M2-branes over $\mathbb{C}^4/\mathbb{Z}_k$ was proposed by Aharony, Bergman, Jafferis and Maldacena (ABJM) \cite{Aharony:2008ug}. The ABJM theory is an $U(N)\times U(N)$ CS gauge theory with levels $k$ and $-k$ and a matter content and superpotential equal to the ones for $N$ D3-branes on the conifold \cite{Klebanov:1998hh}. Soon after the appearance of this model, a lot of activity was devoted to extending these results to cases with reduced SUSY, resulting in the proposal of several gauge theories as candidates for M2-branes over various geometries 
\cite{Benna:2008zy}-\cite{Aganagic:2009zk}.
Several works focused on M2-branes over toric Calabi-Yau 4-folds (CY$_4$) 
\cite{Martelli:2008si}-\cite{Benini:2009qs}.

A remarkable feature of the SCFT on a large number $N$ of M2-branes, which was originally identified in \cite{Klebanov:1996un} from a gravity dual viewpoint, is that its free energy scales as $N^{3/2}$. The attempt to reproduce this scaling from the field theory has been a major driving force for the second front of progress, which concerns the development of methods for counting degrees of freedom in 3d SCFTs (SCFT$_3$). Using localization \cite{Pestun:2007rz}, it has been possible to match the  free energy of the field theory with the dual gravity result for theories with $\mathcal{N} \geq 3$ SUSY \cite{Kapustin:2009kz,Drukker:2010nc,Herzog:2010hf,Gulotta:2011si,Marino:2011eh}. The problem becomes more involved for $\mathcal{N}=2$ theories, for which the free energy is singular. After appropriate regularization, a general expression for the free energy in theories with reduced SUSY was proposed in \cite{Jafferis:2010un,Hama:2010av}. In these cases, the free energy becomes a function of the scaling dimensions (which in 3d are equal to the superconformal R-charges) of fields. Moreover, \cite{Jafferis:2010un} showed that the exact superconformal R-charge is obtained by extremizing the free energy, in the same spirit of a-maximization in 4d \cite{Intriligator:2003jj}. This proposal has been tested both at the perturbative 
\cite{Amariti:2011hw,Niarchos:2011sn,Minwalla:2011ma,Amariti:2011da,Amariti:2011jp}
and non-perturbative levels \cite{Martelli:2011qj,Cheon:2011vi,Jafferis:2011zi}. Actually, in all examples the free energy has been found not only to be extremized but to be maximized. This observation has led 
\cite{Jafferis:2011zi} to conjecture the existence of an $F$-theorem in three dimensions. Several checks of this conjecture have appeared in \cite{Amariti:2011da,Amariti:2011xp,Klebanov:2011gs,Morita:2011cs,Klebanov:2011td}.

Borrowing from the 4d nomenclature, it is useful to distinguish between chiral-like and non-chiral-like theories. As the name indicates, non-chiral-like quivers are those in which every bifundamental field is accompanied by another bifundamental with opposite charges. These techniques have allowed non-trivial checks of the AdS$_4$/CFT$_3$ for non-chiral-like theories \cite{Martelli:2011qj,Cheon:2011vi,Jafferis:2011zi,Amariti:2011uw}. The $N^{3/2}$ scaling of the free energy has not been observed in chiral-like theories yet. This fact might indicate some problem in taking the large-$N$ limit or, more drastically, it can mean that these theories do not describe SCFTs on M2-branes. The answer is still inconclusive, even though some partial results pointing in the first direction have appeared in the literature \cite{Gulotta:2011si,Amariti:2011uw,Gulotta:2011aa,Kim:2012vz}.
One of the main purposes of this paper is, in the spirit of similar calculations for 4d SCFTs (SCFT$_4$) \cite{Benvenuti:2004dy,Benvenuti:2005ja,Franco:2005sm,Butti:2005sw}, to explicitly show the agreement between the field theoretic and gravity determinations of the free energy in infinite classes of models with an arbitrarily large number of gauge groups. In doing so, we accumulate evidence that not only supports the application of the localization ideas to the determination of the free energy in theories with reduced SUSY, but also validates the gauge theories we consider as the correct theories on M2-brane over the corresponding CY$_4$'s.

This paper is organized as follows. In Section \ref{section_background}, we review the computation of the volume of Sasaki-Einstein 7-manifolds at the base of toric CY$_4$ cones, the calculation of the free energy of SCFT$_3$'s, and the correspondence between gauge theory, geometry and dimer models. Section \ref{section_Laba_k} discusses the $L^{a,b,a}_{\vec{k}}$ infinite family of gauge theories, which are the main focus of the paper. These theories have the same quiver and superpotential of $L^{a,b,a}$ models in 4d and \cite{Benvenuti:2005ja,Franco:2005sm,Butti:2005sw}, in addition, CS couplings encoded in $\vec{k}$. Section \ref{section_lifting_algorithm} is devoted to the Type IIB realization of these theories and introduces an algorithm that lifts the cone over $L^{a,b,a}$ to the toric CY$_4$ that corresponds to the mesonic moduli space of the CS quiver. The lifting algorithm is used in Section \ref{section_infinite_families} to generate infinite classes of models, for which the agreement between the geometric and field theoretic determinations of the free energy is established. In Section \ref{section_quartic} we show that, in all the infinite classes of models considered in the paper, it is possible to express the free energy as a quartic function of the R-charges of extremal perfect matchings, even before extremization. We present some thoughts about a general expression for such a quartic function. In Section \ref{section_toric_duality}, we show that the free energy is invariant for certain toric duals obtained by permuting 5-branes in the Type IIB construction of the $L^{a,b,a}_{\vec{k}}$ models. We conclude in Section \ref{section_conclusions}.

\bigskip

\section{Some Background}

\label{section_background}

In this section we review some topics we will later use throughout the paper.

\subsection{Sasaki-Einstein Volumes}

We are interested in the quiver gauge theory on the worldvolume of M2-branes probing a CY$_4$ that is real cone over a 7-dimensional Sasaki-Einstein (SE) manifold $Y_7$. The volume of $Y_7$ is expected to control the number of degrees of freedom of the gauge theory. For toric CY$_4$'s, this volume can be computed from the toric diagram in terms of the Reeb vector $\mathbf{b} = (b_1,b_2,b_3,b_4)$ \cite{Martelli:2005tp}, which is a constant norm Killing vector field commuting with all the isometries of the SE manifold.

There is a one-to-one correspondence between extremal perfect matchings, i.e. corners, of the toric diagram and a basis of 5-cycles $\Sigma_i$ in the base over which M5-branes can be wrapped.\footnote{The concept of perfect matching becomes important when realizing these theories in terms of brane tilings. This is discussed in Section \ref{section_dimers_geometry}. For the purpose of this section, it is sufficient to regard perfect matchings as points in the toric diagram.} The R-charge of a single M5-brane wrapped over $\Sigma_i$  is given by 
\begin{equation}
\Delta_i = \frac{\pi}{6} \frac{\rm{Vol}(\Sigma_i)}{\rm{Vol}(Y_7)} \,.
\end{equation}
This is a function of the Reeb vector $\mathbf{b}$, and the exact superconformal R-charge is obtained by extremizing the function $Z_{\rm MSY}$ defined as
\begin{equation}
Z_{\rm MSY} = \sum_{i=1}^d \rm{Vol}(\Sigma_i) ,
\label{volumeformula}
\end{equation}
where $d$ is the numer of corners of the toric diagram. In terms of $Z_{\rm MSY}$, the volume of $Y_7$ and the R-charges of extremal perfect matchings are
\begin{equation}
{\rm Vol}(Y_7) = \frac{\pi^4}{12} Z_{\rm MSY},
\quad \quad \quad \quad
\Delta_i = \frac{2 {\rm Vol}(\Sigma_i)}{Z_{\rm MSY}} .
\label{Vol_toric_Y7}
\end{equation}
The volumes Vol$(\Sigma_i)$ can be calculated from the toric diagram thanks to the algorithm introduced in \cite{Martelli:2005tp}, extended to CY$_4$'s in \cite{Hanany:2008fj}. Every point in the toric diagram is given by a 4-vector that, due to the Calabi-Yau condition, can be taken to the form $v_i=(\tilde{v}_i,1)$, with $\tilde{v}_i$ a 3-vector. Considering the counterclockwise sequence $w_{k}$, $k=1,\dots,n_i$ of vectors adjacent to a given vector $v_i$ one has
\begin{equation}\label{MSvol}
{\rm Vol}(\Sigma_i) = \sum_{k=2}^{n_{i}-1} \frac{
\langle v_i,w_{k-1},w_k,w_{k+1}\rangle \langle v_i,w_k,w_1,w_{n_i}\rangle}{\langle v_i,b,w_k,w_{k+1}\rangle
\langle v_i,b,w_{k-1},w_k\rangle \langle v_i,b,w_1,w_{n_i}\rangle} ,
\end{equation}
where $\cdot$ indicate column 4-vectors and $\langle \cdot ,\cdot ,\cdot ,\cdot \rangle$ is the determinant of the resulting $4\times 4$ matrix.

\subsection{Free Energy}

\label{section_free_energy}

We now briefly review the calculation of the free energy in 3d, vector-like, CS quivers in the large-$N$ limit. The free energy is computed in terms of the partition function on a 3-sphere $\mathcal{Z}_{S^3}$ as
\begin{equation} \label{FfromZ}
F = -\log |\mathcal{Z}_{S^3}| \,.
\end{equation}
The partition function has been calculated in \cite{Jafferis:2010un,Hama:2010av} by exploiting the localization technique \cite{Pestun:2007rz}, which reduces it to a matrix integral. For a gauge group $G$, with CS level $k$, and matter in the representation ${\bf R}$ of the gauge group with quantum scaling dimension $\Delta$, one has

\begin{equation} \label{Zjafferis}
\mathcal{Z}_{S^3} = \int d \left[\frac{\lambda}{2 \pi} \right] e^{\frac{i k  \text{Tr} \lambda^2}{4 \pi}-   \Delta_m \text{Tr} \lambda}
\, {\det}_{Adj} \left(2 \sinh {\frac{\lambda}{2}} \right) 
{\det}_{{\bf R}} e^{l\left(1-\Delta+ i \frac{\lambda}{2 \pi}\right)} .
\end{equation}
The integral is performed over the Cartan subgroup of the gauge group. The first exponential corresponds to the CS and monopole contributions. The determinants come from the 1-loop contributions of the vector multiplet and the matter fields. For $\mathcal{N}=2$, the 1-loop determinant of matter fields is expressed in terms of the function $l(z)$, which is defined through its derivative as follows

\begin{equation}
l'(z) = - \pi z \cot {\pi z} ,
\end{equation}
and an appropriate normalization.

In this paper, we are interested in computing (\ref{Zjafferis}) in the large-$N$ limit of vector-like quiver gauge theories with gauge group $G=\prod_a U(N)_{k_a}$ and $\sum_a k_a=0$. 

The integral is dominated by the minimum of the free energy. One can distinguish two contributions to the equations of motion, so called long and short range forces. Long range forces cancel in this class of models and only the short range ones contribute \cite{Herzog:2010hf,Martelli:2011qj,Jafferis:2011zi}. The eigenvalue $\lambda_i{(a)}$ of the $a$-th gauge group scale as
\begin{equation} \label{eigenscaling}
\lambda_i^{(a)} = N^{1/2} x_i + i y_i^{(a)}  ,
\end{equation}
where $x$ and $y$ are real \cite{Jafferis:2011zi}. The real part of \eref{eigenscaling} becomes dense, with density $\rho(x)$, while the imaginary part becomes a 
continuous function of $x$, $y_i^{(a)} \rightarrow y_a(x)$.

The free energy follows from the saddle point equations $\partial_\lambda F=0$ \cite{Jafferis:2011zi}. The relevant contributions for the case of vector-like theories with bifundamental and adjoint matter and $\sum_a k_a=0$ are

\begin{eqnarray}
F_{\text{CS}} &=& \frac{N^{3/2}}{2 \pi} \int \rho(x) x \sum_{a=1}^{|G|} k_a y_a dx \nonumber \\
F_{\text{bif}_{ab}} &=& -N^{3/2} \frac{2-\Delta_{ab}^{+}}{2}
\int \rho^2 dx
\left(\left(\delta y_{ab}+\pi \Delta_{ab}^{-}\right)^2 -\frac{\pi^2}{3} \pi^2 \Delta_{ab}^{+} (4-\Delta_{ab}^{+} )\right) \nonumber \\
F_{\text{adj}} &=& \frac{8N^{3/2}}{3}\pi^2 \Delta(1-\Delta) (2-\Delta) \int \rho^2 dx
\label{free_energy_contributions}
\end{eqnarray}
where the first equation is the CS contribution, the second one is the contribution of a bifundamental-antibifundamental pair connecting the $a$-th and the $b$-th nodes, and the last one is the contribution of an adjoint field. We have defined $\Delta_{ab}^{(\pm)}=\Delta_{ab} \pm \Delta_{ba}$. In the partition function one should take into account the diagonal monopole charge, which is given by $\Delta_m=\Delta(T)-\Delta(\tilde{T})$, where $T$ and $\tilde{T}$ are the diagonal monopole and antimonopole operators. Since vector-like models are charge conjugation invariant, $\Delta(T)=\Delta(\tilde{T})$, and we can set $\Delta_m=0$. The bifundamental contribution is only valid when $\delta y_{ab}=y_a-y_b$ is in the regime $|\delta y_{ab} + \pi \Delta_{ab}^{-}|\leq \pi \Delta_{ab}^{+}$.
The leading contribution to the free energy in the large-$N$ limit is then obtained by extremizing the free energy functional over $\rho$ and $y_a$ while imposing the normalization of $\rho$.

As shown in \cite{Kapustin:2009kz,Drukker:2010nc,Jafferis:2011zi}, the supergravity scaling $N^{3/2}$ \cite{Aharony:2008ug} is recovered and the free energy matches the volume computation from AdS/CFT for theories with $\mathcal{N}> 2$ SUSY. The $\mathcal{N}=2$ case is more involved, because R-charges of matter fields usually differ from the classical value $\Delta=1/2$. Indeed, the exact superconformal R-charges is obtained by extremizing the free energy itself \cite{Jafferis:2010un}. 

Some examples of the agreement between the field theory computation of the free energy and the geometric calculation of volumes have been 
presented in \cite{Martelli:2011qj,Cheon:2011vi,   Jafferis:2011zi,Amariti:2011uw,Gulotta:2011aa}. One of the main goals of this paper is to extend this matching to infinite classes of theories with arbitrarily large number of gauge groups, in the spirit of similar tests performed in the context of the AdS$_5$/CFT$_4$ correspondence 
\cite{Benvenuti:2004dy,Benvenuti:2005ja,Franco:2005sm,Butti:2005sw}. Some infinite families of models, consisting of flavored quivers with one or two gauge groups and necklace quivers with $\mathcal{N}\geq 2$ SUSY, have already been considered in the literature \cite{Herzog:2010hf,Gulotta:2011si,Jafferis:2011zi,Gulotta:2011vp}.

The general conjecture is that the free energy of the gauge theory on $S^3$ is related to Vol$(Y_7)$ via
\begin{equation} 
F = N^{3/2} \sqrt{\frac{2 \pi^6}{27 \, {\rm Vol}(Y_7)}}.
\label{F_versus_Vol}
\end{equation} 
We will later see that, in an infinite number of examples, the previous expression holds even off-shell, i.e. even before maximizing the free energy or minimizing the volume.

\subsection{Geometry, Dimer Models and R-charges}

\label{section_dimers_geometry}

In this paper we will focus, as we will discuss in greater detail in Section \ref{section_Laba_k}, on theories with the same quivers of 4d parents and with additional CS couplings for gauge groups. This class of theories can be encoded in terms of brane tilings \cite{Franco:2005sm,Franco:2005rj}, as originally studied in \cite{Hanany:2008cd}.  Their mesonic moduli space is most efficiently described in terms of perfect matchings of the tiling, which are in one-to-one correspondence with the gauged linear sigma model (GLSM) fields in the toric construction of the moduli space, i.e. they map to points in the toric diagram.\footnote{When constructing a toric Calabi-Yau as the moduli space of a gauge theory,  more than one perfect matching might correspond to the same point in the toric diagram.}
The mapping between chiral fields in the quiver $X_i$ and perfect matchings $p_\alpha$ is given by
\beq
X_i = \prod_{\alpha=1}^c p_\alpha^{P_{i\alpha}},
\eeq
where $c$ is the total number of perfect matchings, and $P_{i\alpha}$ is equal to $1$ if the edge in the brane tiling associated to the chiral field $X_i$ is contained in $p_\alpha$ and zero otherwise.

\beq
P_{i\alpha}=\left\{ \begin{array}{ccccc} 1 & \rm{ if } & X_i  & \in & p_\alpha \\
0 & \rm{ if } & X_i  & \notin & p_\alpha
\end{array}\right.
\eeq

A prominent role is played by the subset of {\it extremal} perfect matchings, i.e. those corresponding to {\it corners} of the toric diagram, which we call $\tilde{p}_\mu$, $\mu=1,\ldots,d$. The gauge theory contains a $U(1)_R\times U(1)_F^3 \times U(1)_B^{a+b-2}$ global symmetry group, where $F$ and $B$ indicate flavor and baryonic symmetries, and the extremal perfect matchings are the only ones with non-trivial charges under them \cite{Butti:2005vn}. In other words, the global $U(1)$ symmetries of all chiral fields in the quiver are determined by their $\tilde{p}_\mu$ content. It is then useful to construct a reduced matrix $\tilde{P}$, which is simply a restriction of $P$ to the columns associated with extremal perfect matchings. Its entries are given by  
\beq
\tilde{P}_{i\mu}=\left\{ \begin{array}{ccccc} 1 & \rm{ if } & X_i  & \in & {p}_\mu \\
0 & \rm{ if } & X_i  & \notin & \tilde{p}_\mu
\end{array}\right.
\eeq
Consider any of the global $U(1)$ symmetries, under which $\tilde{p}_u$ has charge $a_\mu$. The charge of a chiral field is then given by

\beq
Q(X_i)=\sum_{\mu=1}^d \tilde{P}_{i\mu} a_\mu .
\label{X_charges_from_pm}
\eeq
In the case of the R-symmetry, the charges of the extremal perfect matchings are constrained by $\sum_{\mu=1}^d a_\mu=2$. For other $U(1)$ symmetries, the constraint is $\sum_{\mu=1}^d a_\mu=0$.

In what follows, we will use these ideas to organize the computation of the free energy, which will involve two steps:

\begin{itemize}

\item  Use \eref{X_charges_from_pm} to parametrize R-charges of matter fields in terms of those of extremal perfect matchings. A corollary of this parametrization is that symmetries of the toric diagram that reduce the number of independent R-charges of extremal perfect matchings also result in a lower dimensional space of R-charges for the R-charges of chiral fields in the quiver. We will exploit this fact in Section \ref{section_infinite_families}.

\item Maximize the free energy over the resulting $(d-1)$-dimensional space. In some cases, we will impose further symmetries to reduce the problem to an extremization over a 1-dimensional space.

\end{itemize}

\smallskip

\section{$L^{a,b,a}_{\vec{k}}$ Theories}

\label{section_Laba_k}

\begin{figure}[h]
\begin{center}
\includegraphics[width=5cm]{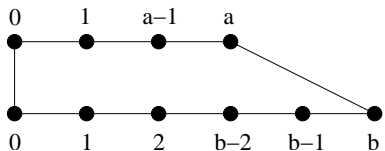}
\caption{Toric diagram for the real cones over $L^{a,b,a}$ manifolds.}
\label{toric_Laba}
\end{center}
\end{figure}

In Section \ref{section_introduction}, we reviewed the extent to which quiver CS theories have been tested as theories on M2-branes and mentioned the difficulties encountered when trying to do so. In order to remain on the conservative side, we will focus in this paper in theories with toric, non-chiral, 4d parents. These parents can be fully classified using toric geometry. They correspond to all toric Calabi-Yau 3-folds (CY$_3$) without compact 4-cycles, i.e. those with toric diagrams without internal points. All geometries satisfying this condition are $\mathbb{C}^{3}/(\mathbb{Z}_2 \times \mathbb{Z}_2)$ and the infinite $L^{a,b,a}$ family. \fref{toric_Laba} shows the toric diagram for the cones over $L^{a,b,a}$ manifolds, consisting of two parallel lines of $(a+1)$ and $(b+1)$ points, respectively. 

\begin{figure}[h]
\begin{center}
\includegraphics[width=13cm]{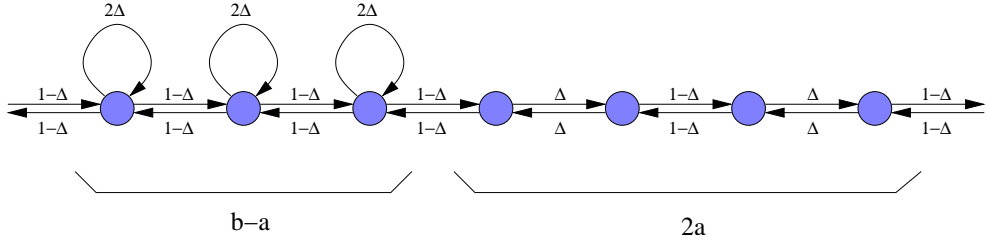}
\caption{Quiver diagram for $L^{a,b,a}$ theories and a one parameter parametrization of the R-charges.}
\label{quiver_Laba}
\end{center}
\end{figure}

The corresponding gauge theory can be taken to the form given in \fref{quiver_Laba} \cite{Benvenuti:2005ja,Franco:2005sm,Butti:2005sw}. The superpotential is given by

\beq
W=\sum_{i=1}^{b-a} X_{i,i}\left(X_{i,i+1}X_{i+1,i} - X_{i,i-1}X_{i-1,i}\right)
+\sum_{i=b-a+1}^{b+a} (-1)^{b+a+i} X_{i,i-1}X_{i-1,i}X_{i,1+1}X_{i+1,i} ,
\eeq
where $X_{i,j}$ indicates a bifundamental field connecting nodes $i$ and $j$ and $X_{i,i}$ corresponds to an adjoint of node $i$. The nodes in the quiver are identified according to $a+b+1 \equiv 1$.

Marginality of the superpotential, which is necessary for conformal invariance, requires that all superpotential terms have R-charge equal to 2. As discussed in the previous section, the number of independent R-charges can be further reduced by the parametrization in terms of extremal perfect matchings in the presence of symmetries of the toric diagram. In the examples studied in Section \ref{section_infinite_families}, symmetries are such that it is possible to express all R-charges in terms of a single parameter $\Delta$ as shown in \fref{quiver_Laba}. This parametrization will be used in Section \ref{section_infinite_families} to deal with some of the more involved examples.

We will add to these models CS couplings for the gauge groups, which can be arranged in a vector $\vec{k}=(k_1,\ldots,k_{a+b})$. We will often use the notation
\beq
\vec{k}=(k_1,\ldots,k_{b-a}||k_{b-a+1},\ldots,k_{a+b}) ,
\label{kvec_notation}
\eeq
where we use a double line to separate nodes with and without an adjoint field. We denote the resulting theories $L^{a,b,a}_{\vec{k}}$. In Section \ref{section_lifting_algorithm}, we introduce an algorithm that determines how the inclusion of $\vec{k}$ lifts the CY$_3$ given by the real cone over $L^{a,b,a}$ to a CY$_4$.

The $\mathcal{N}=3$ necklace quivers of \cite{Herzog:2010hf,Gulotta:2011si,Gulotta:2011vp} have the same matter content of our models for $a=b$, but additional quartic superpotential interactions. $\mathcal{N}=2$ deformations of these theories, obtained by integrating-in adjoint fields and adding polynomial superpotential interactions for them, have also been considered \cite{Jafferis:2011zi}. 

\bigskip

\section{Lifting Calabi-Yau 3-folds to Calabi-Yau 4-folds}

\label{section_lifting_algorithm}

By now, it is well-known that candidates for 3d theories on M2-branes over toric CY$_4$'s can be constructed by starting from theories with the same quivers and superpotentials of 4d theories on D3-branes over toric CY$_3$'s, to which we refer as ``parents", and adding CS terms for the gauge groups. This strategy was exploited soon after the introduction of the ABJM model for generating potential M2-brane theories with reduced SUSY \cite{Imamura:2008nn,Hanany:2008cd,Ueda:2008hx,Franco:2008um,Amariti:2009rb}. The 3d toric diagram of the ``uplifted" CY$_4$ is such that it reduces to the 2d one of the parent CY$_3$ when projected along a direction determined by the CS levels. From the perspective of the computation of moduli spaces, this additional projection arises from an extra D-term constraint that is imposed in the 4d theories. Models in which such a projection is not possible, and hence do not descend from a 4d parent, have also been proposed \cite{Hanany:2008fj,Franco:2008um,Hanany:2008gx,Franco:2009sp,Davey:2009sr}.

\begin{figure}[h]
\begin{center}
\includegraphics[width=11cm]{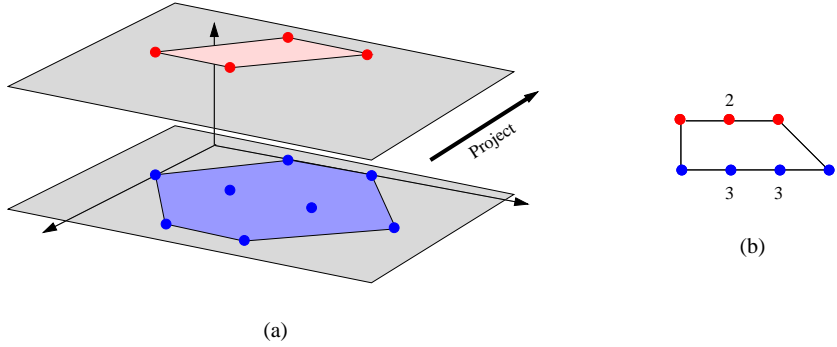}
\caption{A pictorial representation depicting the projection of a toric diagram of a CY$_4$ onto that of a CY$_3$: a) 3d toric diagram of a CY$_4$ and b) its projection onto the $L^{a,b,a}$ geometry.}
\label{3d_projection_toric_Laba}
\end{center}
\end{figure}

In what follows, we will focus our discussion on $L^{a,b,a}_{\vec{k}}$ theories. The most general uplift of \fref{toric_Laba} into a 3d toric diagram corresponds to the two lines turning into convex polygons living on parallel planes, as sketched in \fref{3d_projection_toric_Laba}.

\subsection{A Lifting Algorithm}

In this section we introduce a general algorithm for lifting cones over $L^{a,b,a}$ to CY$_4$'s by appropriate choices of CS levels in the corresponding quivers. We will exploit this procedure in Section \ref{section_infinite_families} for generating interesting classes of models. The method is a specialization of the ideas in \cite{Ueda:2008hx}  to $L^{a,b,a}$ theories.

A useful starting point is the Type IIB brane realization of $L^{a,b,a}_{\vec{k}}$ theories. They can be engineered in terms of an elliptic model consisting of a stack of $N$ D3-branes with one of their worldvolume directions compactified on a circle, suspended between a set of $(b+a)$ $(1,p_i)$ 5-branes. An $(1,p_i)$ 5-brane is a bound state of one NS5-brane and $p_i$ D5-branes. The $p_i$ integers determine the CS levels in the quiver according to the following expression
\beq
k_i = p_{i-1}-p_i  .
\label{CS_levels_from_pi}
\eeq

\begin{figure}
\begin{center}
\includegraphics[width=12cm]{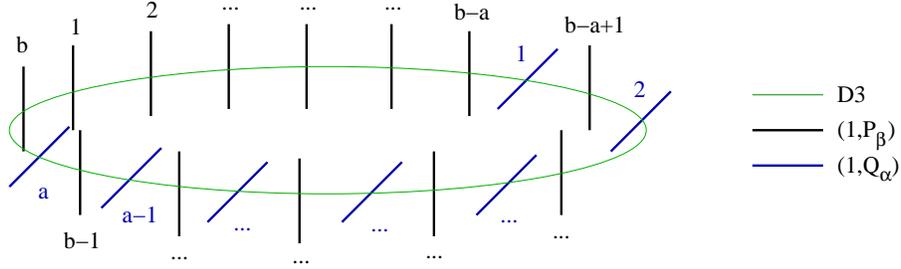}
\end{center}
\caption{Tybe IIB brane system engineering the $L^{a,b,a}_{\vec{k}}$ theories.}
\label{Laba_IIB}
\end{figure}

We split the 5-branes, i.e. the integers $p_i$, into two sets: $Q_\alpha$, $\alpha=1,\ldots,a$, and $P_\beta$, $\beta=1,\ldots,b$.  The branes in the configuration are extended as follows
\beq
\begin{array}{|c|cccccccccc|}
\hline
\ \ {\rm Brane} \ \ & 0&1&2&3& 4&5&6&7&8&9\\
 \hline
{\rm D3} & \times&\times&\times&& &&\times&&&\\
{\rm NS5}_\alpha & \times&\times&\times&\times& \times&\times&&&&\\
{\rm D5}_\alpha & \times&\times&\times&&\times &\times&&\times&&\\
{\rm NS5}_\beta  & \times&\times&\times&\times& &&&&\times&\times\\
{\rm D5}_\beta & \times&\times&\times&& &&&\times&\times&\times\\ \hline
\end{array}
\nonumber
\eeq
The SCFT lives in the $(0,1,2)$ directions common to all the branes. The D3-branes are, in addition, extended along $x_6$, which is compactified on a circle. The $(1,Q_{\alpha})$ 5-brane is a bound state of the NS5$_\alpha$ and $Q_\alpha$ D5$_\alpha$ branes and extends along $(0,1,2,[37]_{\theta_{\alpha}},$ $ 4,5)$. Similarly, the $(1,P_{\beta})$ 5-brane is a bound state of the NS5$_\beta$ and $P_\beta$ D5$_\beta$ branes and extends along $(0,1,2,[37]_{\theta_{\beta}},8,9)$. The final configuration is shown in \fref{Laba_IIB}. In order to reproduce the quiver in 
\fref{quiver_Laba}, we distribute the 5-branes on the circle as follows. First we put $(b-a)$ $(1,P_\beta)$ 5-branes and then we alternate the remaining  $a$ $(1,P_\beta)$ and $a$ $(1,Q_\alpha)$ branes. It is possible to reorder the 5-branes along the $x_6$ circle, which results in dual gauge theories. 

\begin{figure}[h]
\begin{center}
\includegraphics[width=13cm]{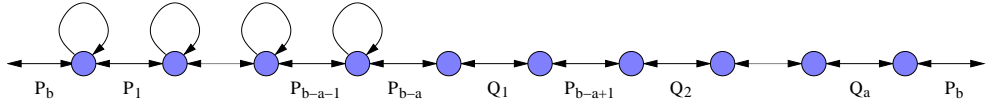}
\caption{Quiver diagram for $L^{a,b,a}$ theories showing the $P_\beta$ and $Q_{\alpha}$ charges.}
\label{quiver_Laba_PQ}
\end{center}
\end{figure}

The D3-branes stretched between each pair of 5-branes gives rise to a gauge group in the quiver. Each 5-brane is associated to a pair of bifundamental chiral fields as shown in \fref{quiver_Laba_PQ}. In addition, we have an adjoint chiral field for each consecutive pair of 5-branes of the same type.

The toric diagram for $L^{a,b,a}$ theories was given in \fref{toric_Laba}. 
As we will now explain, the $P_\beta$'s control how the top line 
of the toric diagram is lifted to a plane. Similarly, 
the $Q_\alpha$'s determine the lift of the bottom line. The degeneracies of perfect matchings associated to points in the toric diagram are $\binom{b}{\mu}$ and $\binom{a}{\nu}$, where $\mu=0,\dots,b$ and $\nu=0,\dots,a$ run over the 
points on the bottom and top row respectively.

Perfect matchings correspond to certain collections of edges in the 
associated brane tilings, which map to sets of chiral fields in the quiver. 
Indeed, thinking in terms of the quiver provides a clear visualization of 
these multiplicities. Let us first consider the $(b+1)$ points in the lowest 
line of the toric diagram. The perfect matching for $\mu=0$ consists 
of all the fields in the quiver  with the arrows pointing from right to left. The perfect matchings for the $\mu$-th point correspond to reversing the 
orientation of $\mu$ of the fields, giving rise to the multiplicity described
 by the binomial coefficients. Repeating this procedure, we reach the 
 $\mu=b$ point in which all the fields are arrows in the quiver point from left to right.
The line with $(a+1)$ is constructed in the same way, by using the 
fields labeled by $Q_\alpha$, but also including the adjoint fields.

The new mesonic direction in the CY$_4$ is determined by the $P_\beta$ or $Q_\alpha$ charges. We set the CS ``fluxes" such that every bifundamental field pointing from left to right carries a flux $P_\beta$ or $Q_\alpha$, while fields from right to left carry zero flux. The new mesonic direction in the CY$_4$ is determined by the $P_\beta$ or $Q_\alpha$ charges following a simple prescription:

\bigskip

\begin{center}

\begin{tabular}{|l|}
\hline
Every perfect matching in the 2d toric diagram gets a shift into the third \\ dimension equal to the total CS flux it carries. \\
\hline
\end{tabular}

\end{center}

\bigskip

Let us first consider the effect of this rule on the lowest line.
The first point, $\mu=0$, does not have any flux and hence does not move. The other endpoint of the line, $\mu=b$, gets the maximum possible shift, equal $\sum_{\beta=1}^{b} P_\beta$. The intermediate points are not only shifted but they can also be split, depending on the total flux of each of the perfect matchings associated to a given point. The expansion of the top line of the toric diagram into the third dimension follows the same prescription, with fluxes determined by $Q_\alpha$. Positivity of the $P_\beta$ and $Q_\alpha$ charges guarantees the convexity of the resulting 3d toric diagram.

It is possible to take the theory to a conventional form in which the $P_\beta$ are arranged in increasing order
\begin{equation}
P_1 \leq P_2 \leq \dots \leq P_{b-1} \leq  P_b \, ,
\label{sorted_P}
\end{equation}
and similarly for the $Q_\alpha$. In order for all perfect matchings to get different shifts, they must have different fluxes. Then, a necessary condition for fully lifting the degeneracy of points in the bottom and top lines of the toric diagram is that all $P_\beta$ and ll $Q_\alpha$ are different, respectively.

The algorithm we have just described leads to a broad range of results. For example, in the simple case in which $P_\beta\equiv P$ for all $\beta$ and $Q_\alpha=Q$  (with $Q \neq P$) for all $\alpha$, the two lines are lifted to lines, giving rise to the toric diagram for $\mathbb{C}^2/\mathbb{Z}_a\times \mathbb{C}^2/\mathbb{Z}_b$. This is the situation considered in \cite{Imamura:2008nn}. On the other end of the spectrum, we have cases in which every internal point of the lines is expanded and generates two new corners. Together with the four external points of the original 2d toric diagram, they lead to a toric diagram with $2(a+b)$ corners. A necessary condition for this to happen is that all the inequalities in \eref{sorted_P} are strict.

Let us now discuss in further detail the lift of internal points inside the two lines in the $L^{a,b,a}$ toric diagram. We discuss the bottom line, the other one behaves in a similar way. The $\mu=1$ point expands into a segment in which, if we sort the $P_\beta$'s as in \eref{sorted_P}, the bottom and top endpoints are shifted by $P_1$ and $P_b$ units, respectively. I.e., this point turns into segment of length $(P_b-P_1)$.\footnote{Clearly, if all $P_\beta$ are equal, the segment degenerates into a point.} The $\mu=b-1$ point also expands into a segment, with its endpoints shifted by $\sum_{\beta=1}^{b-1} P_\beta$ and $\sum_{\beta=2}^{b} P_\beta$. Once again, the length of the resulting segment is $(P_b-P_1)$. The same phenomenon occurs for all other internal point, i.e. the $\mu$-th and $(b-\mu)$-th points turn into segments of equal length. The maximal length is attained for the $(b-1)/2$-th and the $(b+1)/2$-th points for odd $b$, and for the $b$-th point for odd $b$.

\subsection*{An Example}

\begin{figure}[h]
\begin{center}
\includegraphics[width=8.5cm]{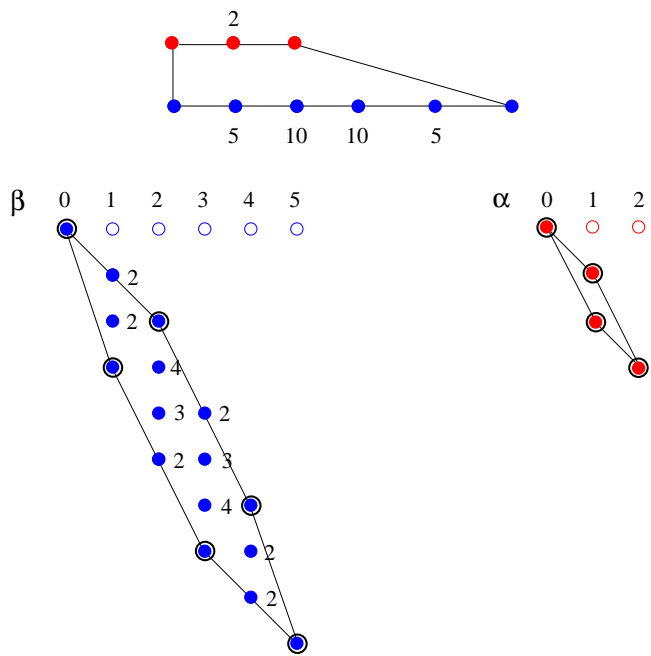}
\caption{Lift of the toric diagram of $L^{2,5,2}$ for $P_\beta = \{1,2,3,1,2 \}$ and $Q_\alpha = \{2,1 \}$. We indicate the multiplicity of internal points and identify the corners of the 3d toric diagram with black circles.}
\label{toric_L252}
\end{center}
\end{figure}

Let us illustrate the previous ideas with an explicit example. Consider the $L^{2,5,2}$ theory and take

\beq
\begin{array}{ccl}
P_\beta & = & \{1,2,3,1,2 \} \\
Q_\alpha & = & \{2,1 \}
\end{array}
\eeq
which, following \eref{CS_levels_from_pi}, generates the following CS levels for the quiver
\beq
k_i  = \{1,-1,-1,1,1,0,-1 \}  .
\eeq
\fref{toric_L252} shows the result of applying the lifting algorithm. We see the, in this case partial, lift of degeneracies of points in the toric diagram and the appearance of new corners.

\bigskip

\section{Infinite Families}

\label{section_infinite_families}

In this section we present various infinite families of gauge theories and the associated CY$_4$'s obtained from $L^{a,b,a}$ models by the lifting algorithm introduced in Section 
\ref{section_lifting_algorithm}. In all these cases we show the volume computation and the gauge theory calculation of the free energy agree. Interestingly, this agreement holds even off-shell.

We present geometries whose toric diagrams have 4, 6 and 8 extremal perfect matchings. The latter models are interesting because they give rise to non-trivial R-charges.

\subsection{Four extremal points: $L^{a,b,a}_{(0,\dots,0||k,-k,\dots,k,-k)}$} 

\label{section_4_points_family_1}

\begin{figure}[h]
\begin{center}
\includegraphics[width=6.5cm]{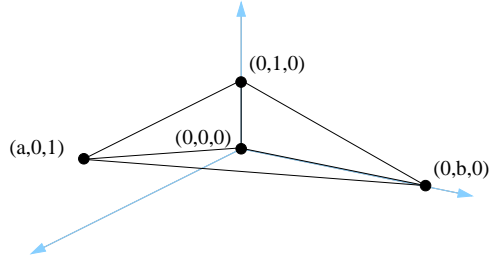}
\caption{Toric diagram for the $L^{a,b,a}_{(0,\dots,0||k,-k,\dots,k,-k)}$ family with $k=1$.}
\label{toric_4_corners}
\end{center}
\end{figure}

We start our investigation of infinite classes of models by considering geometries whose toric diagrams, shown in \fref{toric_4_corners}, have four corners given by the vectors

\begin{equation}
\left(
\begin{array}{cccccc}
\ \ v_1 \ \ & \ \ v_2 \ \ & \ \ v_3 \ \ & \ \ v_4 \ \ \\
 0 & 0 & 0 & a \\
 0 & b & 0 & 0 \\
 0 & 0 & 1 & 1 \\
 1 & 1 & 1 & 1  
\end{array}
\right)
\label{matrix_toric_4_corners}
\end{equation}

These geometries are $\mathbb{C}^2/\mathbb{Z}_a \times \mathbb{C}^2/\mathbb{Z}_b$ orbifolds and their dual gauge theories were introduced and investigated in \cite{Imamura:2008nn}. They are obtained via the lifting algorithm by setting, for example, 

\beq
P_\beta=k, \qquad \qquad Q_\alpha=0.
\eeq
The resulting CS couplings are

\beq
\vec{k}=(0,\dots,0||k,-k,\dots,k,-k).
\eeq

\subsection*{Geometric computation}

The $Z_{\rm MSY}$ function is obtained by summing over the volumes of the 5-cycles corresponding to the extremal points in the toric diagram.
They are functions of the Reeb vector $\mathbf{b}$ and $Z_{\rm MSY}$, which 
corresponds to the sum of these contribution, becomes
\begin{equation}
Z_{\rm MSY}=
\frac{4 a b}{b_1 b_2 \left(b_2+b \left(b_3-4\right)\right) \left(b_1-a b_3\right)},
\end{equation}
where we have set, as in all the examples that follow, $b_4=4$.
Assigning an R-charge $\Delta_i$ to each of the four extremal points $v_i$,
these $R$-charges, corresponding to the charges of the extremal perfect matchings,
can be expressed in terms of the Reeb vector as
\begin{equation}
\Delta_1= -\frac{b_2+b \left(b_3-4\right)}{2 b},\quad
\Delta_2=\frac{b_2}{2 b},\quad
\Delta_3=-\frac{b_1-a b_3}{2 a},\quad
\Delta_4=\frac{b_1}{2 a} ,
\end{equation}
and the $Z_{\rm MSY}$ function becomes
\begin{equation} \label{Volaba}
\text{Vol}(Y_7) = \frac{\pi^4}{48 a b k \,\Delta_1  \Delta_2  \Delta_3  \Delta_4}.
\end{equation}
Under the constraint $\sum_i \Delta_i=2$, the volume is minimized for $\Delta_i=1/2$. We have included an extra $k$ factor with respect to \eref{Vol_toric_Y7} in the denominator of the volume due to an additional $\mathbb{Z}_k$ orbifold action on the moduli space, where $k=\text{gcd} (\{k_a\})$ \cite{Aharony:2008ug}. This factor will also be present in the volumes of all the examples that follow.

\subsection*{Free energy computation}

\begin{figure}[h]
\begin{center}
\includegraphics[width=11cm]{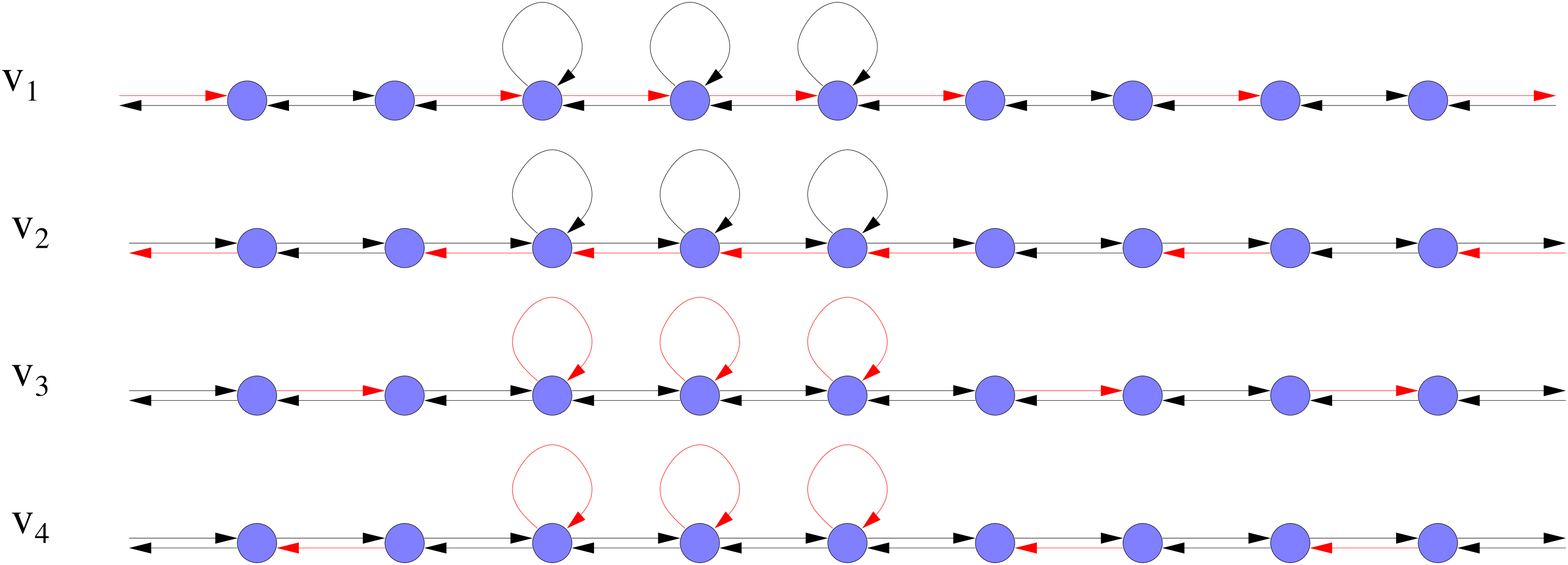}
\caption{Perfect matchings associated to the four corners of the toric diagram given in \eref{matrix_toric_4_corners}. 
Red arrows indicate the chiral fields associated with edges in the perfect matching.}
\label{PMLaba}
\end{center}
\end{figure}

Let us now compute the free energy of this class of models. Recall that a perfect matching is a subset of edges such that every vertex in the brane tiling is an endpoint of precisely one edge in the set. Using the dictionary between brane tilings and gauge theories \cite{Franco:2005rj}, a perfect matching can be interpreted as a subset of the chiral fields in the quiver such that it contains exactly one field for each superpotential term. The four extremal perfect matchings can be simply represented in terms of the quiver as shown in \fref{PMLaba}.

\begin{figure}[h]
\begin{center}
\includegraphics[width=14cm]{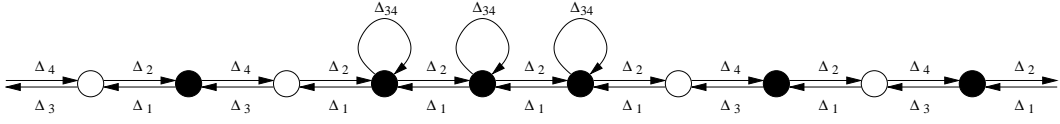}
\caption{R-charges of chiral fields in terms of the R-charges of extremal perfect matchings. We have defined $\Delta_{34} \equiv \Delta_3+\Delta_4$.}
\label{Quiver_Laba_BW}
\end{center}
\end{figure}

It is then straightforward to determine the matrix 
$\widetilde P_{i\mu}$
and the R-charges of chiral fields in terms of those of the extremal perfect matchings. We show the result in \fref{Quiver_Laba_BW}.

The free energy is given by the sum of the CS and the matter field
(bifundamentals and adjoints) contributions. 
As we already anticipated we are setting the monopole charge to zero
even off-shell.
The CS contribution to the large-$N$ free energy is
\begin{equation}
\frac{F_{\text{CS}}}{N^{3/2}}= \sum_{i=1}^{a} \frac{k}{2 \pi}\int{\rho\, \delta y_{b-a+2i-1,b-a+2i} \,x \, dx}.
\end{equation}

The matter contribution is
\begin{eqnarray} \label{matterLaba}
\frac{F_\text{matter}}{N^{3/2}}= &-&\sum_{i \in e(B)} \frac{2-\Delta_{i,i+1}^{(+)}}{2}  \int{\rho^2 
\left(\left(\delta y_{i,i+1} +\pi \Delta_{i,i+1}^{(-)}\right)^2
-
\frac{\pi^2}{3} \Delta_{i,i+1}^{(+)}
 \left(4-\Delta_{i,i+1}^{(+)}\right)
\right) 
dx}
\nonumber \\
&-&\sum_{i \in e(W)} \frac{2-\Delta_{i,i+1}^{(+)}}{2}  \int{\rho^2 
\left(\left(\delta y_{i,i+1} +\pi \Delta_{i,i+1}^{(-)}\right)^2
-
\frac{\pi^2}{3} \Delta_{i,i+1}^{(+)}
 \left(4-\Delta_{i,i+1}^{(+)}\right)
\right)
dx}
\nonumber \\
&+&\frac{2 \pi^2}{3} \sum_{i\in e(B)'}
 \Delta_{i,i+1}^{(+)} \left(1-\Delta_{i,i+1}^{(+)}\right) \left(2-\Delta_{i,i+1}^{(+)}\right)
\int \rho^2 dx ,
\end{eqnarray}
where $e(B)$ and $e(W)$ refer to the black and white nodes in the quiver as shown in \fref{Quiver_Laba_BW}. We moreover denote $e(B)'$ the subset of the black nodes containing adjoint fields. We have defined $\Delta_{i,j}^{(\pm)}=\Delta_{i,j}+\Delta_{j,i}$. We solve the Euler-Lagrange equations with the $\delta y$ variables subject to the following constraint
\begin{equation}
\delta y_{b+a,1}= \sum_{i=1}^{b+a-1} \delta y_{i,i+1} .
\end{equation}

R-charges are parametrized as in \fref{Quiver_Laba_BW}. In the sum over black nodes, we can rewrite $\Delta_{i,i+1}^{(+)}= \Delta_1+\Delta_2 = \Delta_{12}$ and $\Delta_{i,i+1}^{(-)}= \Delta_1-\Delta_2$.
Similarly, in the sum over white nodes we can rewrite 
$\Delta_{i,i+1}^{(+)}= \Delta_3+\Delta_4 = \Delta_{34}$ and 
$\Delta_{i,i+1}^{(-)}= \Delta_3-\Delta_4$.
The eigenvalue distribution is reduced to a piecewise function over three connected domains as follows

\begin{equation}
\left \{
\begin{array}{rcl}
 \delta  y_W & = & -{b\over a}\delta y_B=-
\frac{4 b k \pi ^2 x \left(b \Delta_4\Delta_3\Delta_{12}+a \Delta_1\Delta_2\Delta _{34}\right)+2 b \pi  \mu  \left(\Delta_2 \Delta_3-\Delta_1 \Delta_4+\Delta_2 \Delta_4\right)}{2 a b k \pi  x \left(\Delta_2 \Delta_3-\Delta_1 \Delta_4+\Delta_3 \Delta_4\right)-\mu  \left(a \Delta_{12}+b \Delta _{34}\right)} 
\\ 
& & \qquad \qquad \qquad \qquad \qquad \qquad \qquad \qquad \qquad \qquad \qquad \qquad -\frac{\mu }{2 b k \pi  \Delta_4}<x<\frac{\mu}{2 b k \pi  \Delta_3} 
\\
\delta \rho & = & \frac{\mu  \left(a \Delta _{12}+b \Delta_{34}\right)-2 a b k \pi  x \left(\Delta_2 \Delta_3-\Delta_1 \Delta_4+\Delta_2 \Delta_4\right)}{8 \pi ^3 \Delta_{12} 
\Delta_{34} \left(a \Delta_1+b \Delta_3\right) \left(a \Delta_2+b \Delta_4\right)} 
\end{array}
\right.
\end{equation}
Out of this region, we have
\begin{equation}
\left \{
\begin{array}{rclcc}
\delta y_W & = & -\frac{b}{a}\delta y_B = -2 \pi \Delta_1 
& \qquad \qquad \qquad &
-\frac{\mu}{2 b k \pi  \Delta_1}<x<-\frac{\mu }{2 b k \pi  \Delta_4} \\ \\
\rho & = & -\frac{b \left(\mu +2 a k \pi  x \Delta_1\right)}{8 \pi ^3 \Delta_{12} \left(a \Delta_1+b \Delta_3\right) \left(a \Delta_1-b \Delta_4\right)} & &
\end{array}
\right.
\end{equation}
and
\begin{equation}
\left \{
\begin{array}{rclcc}
\delta y_W & = & -\frac{b}{a}\delta y_B = 2 \pi \Delta_3 
& \qquad \qquad \qquad &
\frac{\mu }{2 b k \pi  \Delta_3} <x<\frac{\mu}{2 a k \pi  \Delta_2}
\\ \\
\rho & = & -\frac{b \left(\mu -2 a k \pi  x \Delta_2\right)}{8 \pi ^3 \Delta_{12} \left(a \Delta_2-b \Delta _3\right) \left(a \Delta_2+b \Delta_4\right)} & &
\end{array}
\right.
\end{equation}
Integrating over the piecewise domain and imposing the
normalization on $\rho$ we obtain
\begin{equation}
\frac{F^2}{N^{3}} = 
\frac{32}{9} a b k \pi ^2 \Delta_1\Delta_2\Delta_3\Delta_4,
\end{equation}
in perfect agreement with \eref{Volaba} via \eref{F_versus_Vol}.

\bigskip

\subsection{Six extremal points} 

\subsubsection{Family 1: $L^{a,b,a}_{(k,0,\dots,0||-k,0,\dots,0)}$}

\label{section_6_points_family_1}

We consider a family with the toric diagram shown in \fref{matrix_toric_6_corners_1}, whose corners are given by the following vectors

\begin{equation} 
\left(
\begin{array}{cccccc}
\ \ v_1 \ \ & \ \ v_2 \ \ & \ \ v_3 \ \ & \ \ v_4 \ \ & \ \ v_5 \ \ & \ \ v_6 \ \  \\
 0 & 0 & a & a & 0 & a \\
 0 & 0 & 0 & 0 & 1 & 1  \\
 0 & b-a & 0 & b-a & 0 & 0  \\
 1 & 1 & 1 & 1 & 1 & 1 
\end{array}
\right)
\label{matrix_toric_6_corners_1}
\end{equation}

\begin{figure}[h]
\begin{center}
\includegraphics[width=5cm]{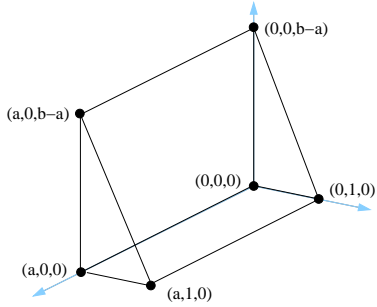}
\caption{Toric diagram for the $L^{a,b,a}_{(k,0,\dots,0||-k,0,\dots,0)}$ family with $k=1$.}
\label{toric_6_corners_1}
\end{center}
\end{figure}

The charges associated to the lifting algorithm are

\begin{equation} \label{table_PQ_Laba_6points_1}
\begin{array}{|c|cc|}
\hline 
\ {\rm Type} \ & \ {\rm Multiplicity} \ & \ {\rm Value} \ \\
 \hline
P & b-a & 0\\
P & a & k \\
Q & a & k\\
\hline
\end{array}
\end{equation}
which results in the following CS levels
\beq
\vec{k}=(k,0,\dots,0||-k,0,\dots,0).
\eeq
This family contains and generalizes the D3 model considered in \cite{Hanany:2008fj,Franco:2008um}, which corresponds to $L^{1,2,1}_{(k||-k,0)}$.

\subsection*{Geometric computation}

$Z_\text{MSY}$ can be written as
\begin{equation}
Z_\text{MSY}=
\frac{16 a (b-a)}{\left(4 a-b_1\right)\left((b-a) \left(4-b_2\right)+b_3\right) b_1 b_2 b_3} .
\end{equation}
Exploiting the symmetries of the toric diagram,\footnote{In this example and the ones that follow, it is possible to make the symmetries of the toric diagram more manifest by acting with appropriate $SL(3,\mathbb{Z})$ transformations.} which identifies the R-charges of certain perfect matchings, we can parametrize the components of the Reeb vector as
\begin{equation}
b_1 = 2 a,  \quad b_2 = 4 \Delta, \quad b_3 = 2(b-a)(1-\Delta)
\end{equation}
and the volume becomes
\begin{equation}
\text{Vol}(Y_7)=
\frac{\pi ^4}{48 k\, a (b-a) (1-\Delta )^2 \Delta} .
\label{Vol_6_corners_1}
\end{equation}
The volume is minimized for $\Delta=1/3$.

\subsection*{Free energy computation}

\begin{figure}[h]
\begin{center}
\includegraphics[width=11cm]{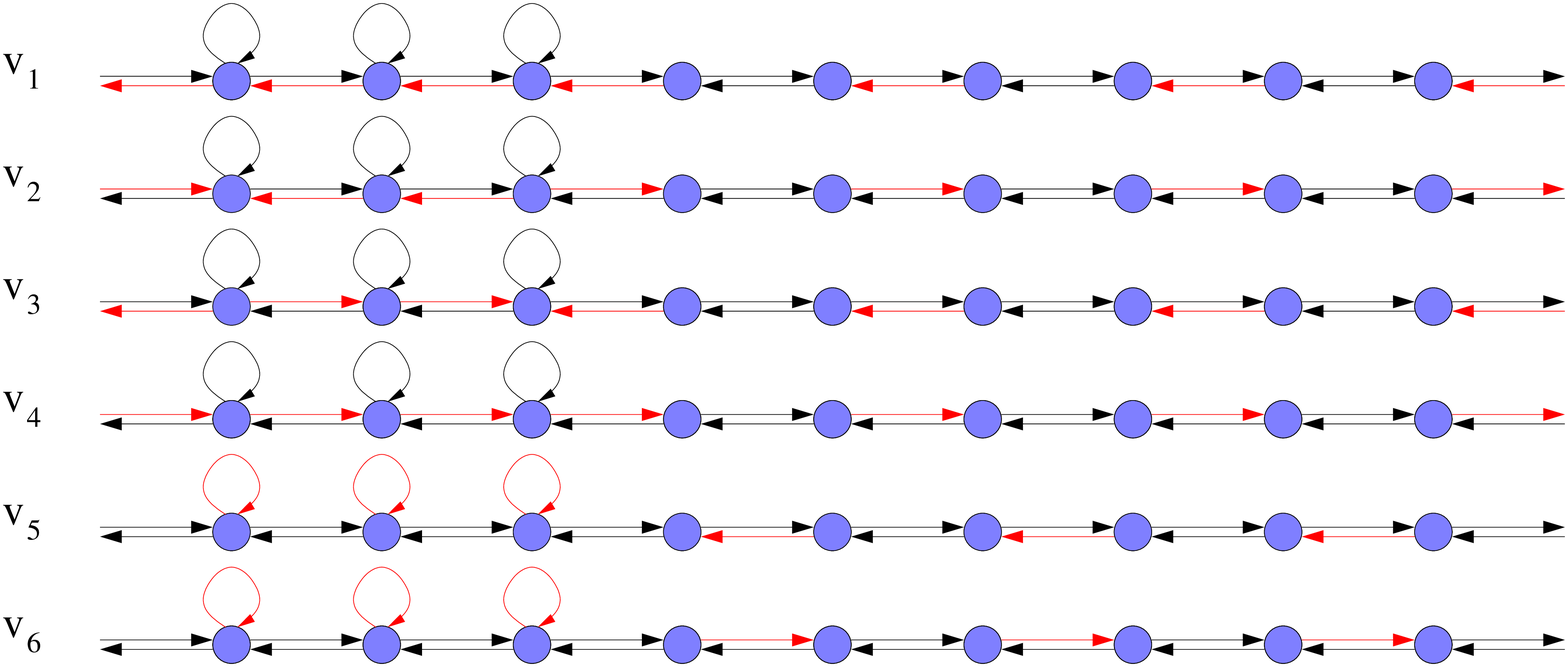}
\caption{Perfect matchings associated to the six corners of the toric diagram given in 
\eref{matrix_toric_6_corners_1}. 
Red arrows indicate the chiral fields associated with edges in the perfect matching.}
\label{PM_Laba_6_points_1}
\end{center}
\end{figure}

Before computing the free energy, we specify the 
perfect matchings  as collections of chiral fields in the quiver. The six 
extremal perfect matchings for this class of models are represented in terms of the quiver as shown in \fref{PM_Laba_6_points_1}.

\begin{figure}[h]
\begin{center}
\includegraphics[width=14cm]{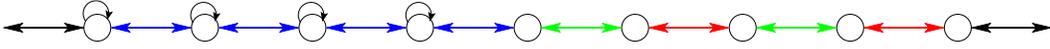}
\caption{Different sets of the imaginary parts of the eigenvalues.}
\label{deltayLaba}
\end{center}
\end{figure}

The CS contribution to the free energy in this case is
\begin{equation}
\frac{F_{\text{CS}}}{N^{3/2}}= \sum_{i=1}^{b-a} \frac{k}{2 \pi}\int{\rho \delta y_{i,i+1} x dx} .
\end{equation}
The sum over matter fields can be organized as follows. First, we distinguish three different kinds
of $\delta y$'s: ``red", ``green" and ``blue" as in \fref{deltayLaba}.
%
%
%
%
%
%
%
%
Notice that we enforced the constraint $\sum \delta y=0$
by drawing $\delta y_{b+a,1}$ in black.
Moreover one can check that the equations of motion give the same value to the 
$\delta y$'s with the same color. Using the marginality of the superpotential and symmetries coming from the toric diagram, we can parametrize the R-charges in terms of a single parameter $\Delta$ as explained in Section \ref{section_Laba_k}. We can then write the matter contribution to the free energy as
{\small
\begin{eqnarray}
\frac{F_{\text{matter}}}{N^{3/2}} & = & 
- \Delta \int \rho^2 \left(\left((b-a) \delta y_b+(a-1) \delta y_r+a \delta y_g
 \right)^2 - \frac{4}{3} \pi^2 \left(1-\Delta^2\right)\right) dx\\
& - & (b-a) \Delta \int \rho^2 \left(\delta y_b - \frac{4}{3} \pi^2 \left(1-\Delta^2\right)\right)dx
-(a-1) \Delta \int \rho^2\left(\delta y_r - \frac{4}{3} \pi^2 \left(1-\Delta^2\right) \right)dx
\nonumber\\& - &
a (1-\Delta) \int \rho^2 \left(\delta y_g - \frac{4}{3} \pi^2\Delta \left(2-\Delta\right)\right)dx 
+
\frac{8}{3} (b-a) \pi ^2 (1-\Delta ) \Delta  (1-2 \Delta )\int \rho ^2 dx
\nonumber
\end{eqnarray}}
The Euler-Lagrange equations give

\begin{equation}
 \left \{
 \begin{array}{lcrcccl}
  \rho = \frac{a (2 (b-a) k \pi  x (1-\Delta )+\mu )}{16 \pi ^3 (1-\Delta ) \Delta  (a (2-\Delta )-b(1- \Delta )) (b(1-\Delta )+a \Delta )} 
 & \ \ \ &
 -\frac{\mu }{2 (b-a) k \pi  (1-\Delta )} & < & x & < & -\frac{\mu }{2 a k \pi }
 \\
 \rho= \frac{\mu }{16 \pi ^3 (1-\Delta ) \Delta  (b (1-\Delta )+a \Delta )}
 & &
 -\frac{\mu }{2 a k \pi } & < & x & < & \frac{\mu }{2 a k \pi } 
\\
 \rho=\frac{a (2 (b-a) k \pi  x (1-\Delta )-\mu )}{16 \pi ^3 (1-\Delta ) \Delta  (b(1-\Delta )-a (2-\Delta )) (b+a \Delta -b \Delta )}
 & &
 \frac{\mu }{2 a k \pi } & < & x & < & 
 \frac{\mu }{2 (b-a) k \pi  (1-\Delta )}
  \end{array}
  \right.
 \end{equation}
Integrating this distribution, we have
\begin{equation}
\frac{F^2}{N^{3}} = 
\frac{32}{9} \pi ^2 k \,a (b-a)   (1-\Delta )^2 \Delta, 
\end{equation}
in agreement with the geometric computation \eref{Vol_6_corners_1}. 
As can be easily observed in (\ref{table_PQ_Laba_6points_1})
$b$ has to be greater than $a$ otherwise all the CS levels 
vanish and the model is not associated to a SCFT in 3d.

\subsubsection{Family 2: $L^{a,2a,a}_{(0,\dots,0,-2k||k,k,-k,k,-k,\dots,k,-k,k)}$}

\label{section_6_points_family_2}

\begin{figure}[h]
\begin{center}
\includegraphics[width=5.5cm]{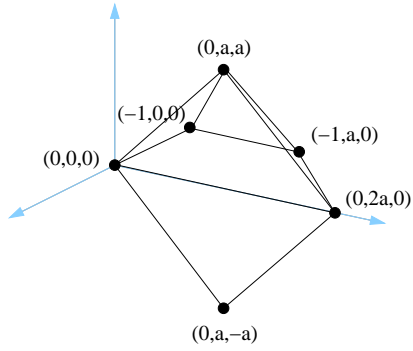}
\caption{Toric diagram for the $L^{a,2a,a}_{(0,\dots,0,-2k||k,k,-k,k,-k,\dots,k,-k,k)}$ family with $k=1$.}
\label{toric_6_corners_2}
\end{center}
\end{figure}

We now consider models with toric diagram given in \fref{toric_6_corners_2}. The six corners of the toric diagram have coordinates given by the matrix

\begin{equation}
\left(
\begin{array}{cccccc}
\ \ v_1 \ \ & \ \ v_2 \ \ & \ \ v_3 \ \ & \ \ v_4 \ \ & \ \ v_5 \ \ & \ \ v_6 \ \  \\
 0 & -1 & -1 & 0 & 0 & 0 \\
 2a & a & 0 & a & a & 0  \\
 0 & 0 & 0 & a & -a & 0  \\
 1 & 1 & 1 & 1 & 1 & 1 
\end{array}
\right)
\label{matrix_toric_6_corners_2}
\end{equation}

These models are constructed by setting $b=2a$ and applying the lifting algorithm with

\begin{equation}
\begin{array}{|c|cc|}
\hline 
\ {\rm Type} \ & \ {\rm Multiplicity} \ & \ {\rm Value} \ \\
 \hline
P & a & 0 \\
P & a & 2k \\
Q & a & k \\
\hline
\end{array}
\end{equation}
The CS levels are
\beq
\vec{k}=(0,\dots,0,-2k||k,k,-k,k,-k,\dots,k,-k,k).
\eeq

This class of theories contains and generalizes the modified SPP model studied in \cite{Hanany:2008fj}, which corresponds in this notation to $L^{1,2,1}_{(-2||1,1)}$, for which the agreement between the free energy and the volume has been shown in \cite{Martelli:2011qj}.

\subsection*{Geometric computation}

The volumes are written in terms of the components of the Reeb vector and we have
\begin{equation}
Z_{\rm MSY}=
\frac{8 a^2 \left(256 a^2-16 b_3^2+b_1 \left(128 a^2+a^2 b_1 \left(20+b_1\right)-a \left(8+b_1\right) b_2+b_2^2-3 b_3^2\right)\right)}{b_1 \left(a^2 \left(4+b_1\right){}^2-b_3^2\right) \left(b_2^2-b_3^2\right) \left(\left(-a \left(8+b_1\right)+b_2\right){}^2-b_3^2\right)}.
\end{equation}
Using the symmetry of the toric diagram we can set $b_3=0$. By applying an $SL(4,\mathbb{Z})$ transformation one can see that $\Delta_2=\Delta_3$. By imposing this symmetry on the components of the Reeb vector we have
\begin{equation}
b_2= \frac{a}{2} \left(b_1+8\right).
\end{equation}
For $\Delta_2=\Delta$ the R-charges of the extremal perfect matchings can be parametrized as
\begin{equation} \label{charge geometry}
\Delta _1=\Delta _6=\frac{2 (\Delta -1)^2}{4-3 \Delta },\quad
\Delta _2=\Delta _3=\Delta , \quad
\Delta _4=\Delta _5=\frac{(\Delta -2) (\Delta -1)}{4-3 \Delta }
\end{equation}
and the volume with this parametrization becomes
\begin{equation}
\text{Vol}(Y_7)=\frac{\pi ^4 (4-3 \Delta )}{96 a^2 k \Delta (\Delta -1)^2(\Delta -2)^2} ,
\end{equation}
which is minimized for
\begin{equation}
\Delta=\frac{1}{18} \left(19-\frac{37}{\left(431-18 \sqrt{417}\right)^{1/3}}-\left(431-18 \sqrt{417}\right)^{1/3}\right) .
\end{equation}
We see that this infinite family of theories already generates rather non-trivial R-charges. This is also the case for the families with eight corners in the toric diagram discussed in the next section.

\subsection*{Free energy computation}

\begin{figure}[h]
\begin{center}
\includegraphics[width=11cm]{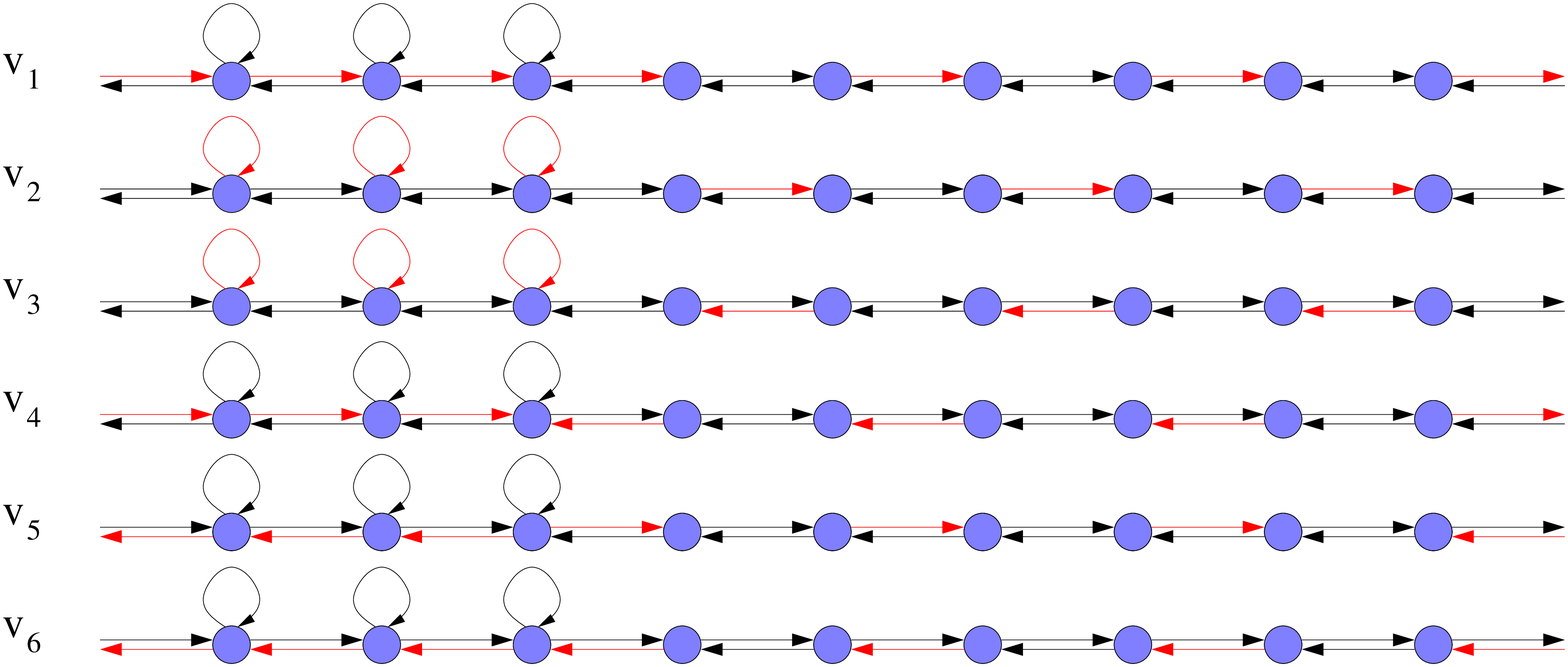}
\caption{Perfect matchings associated to the six corners of the toric diagram given in \eref{matrix_toric_6_corners_2}. Red arrows indicate the chiral fields associated with edges in the perfect matching.}
\label{PM_Laba_6_points_2}
\end{center}
\end{figure}

On the field theory side the six extremal perfect matchings are associated to chiral fields as in \fref{PM_Laba_6_points_2}. The geometric parametrization of R-charges in \eref{charge geometry} then results in the R-charges obtained from imposing marginality of the superpotential and symmetries, which are shown in \fref{quiver_Laba}.

%
%
%
%
After we parametrize the charges as in the geometrical side we can calculate the large-$N$ free energy and compare it to the volume.
The large-$N$ free energy has the following contribution from the CS term
\begin{equation}
\frac{F_{\text{CS}}}{N^{3/2}}= \sum_{i=1}^{a} \frac{k}{2 \pi}\int{\rho \left(2\delta y_{a+2i-1,a+2i-2}+
\delta y_{a+2i,a+2i-1}\right) x \, dx},
\end{equation}
while matter fields give
{\footnotesize
\begin{eqnarray}
\frac{F_{\text{matter}}}{N^{3/2}} = &&
(\Delta-1) \sum _{i \in e(B)}  \int \rho ^2 \left(\text{$\delta $y}_i^2-\frac{4}{3} \pi ^2 (2-\Delta ) \Delta \right)dx
-
\Delta \sum _{i\in e(W)}  \int \rho ^2 \left(\text{$\delta $y}_i^2-\frac{4}{3} \pi ^2 \left(1-\Delta ^2\right) \right)dx
\nonumber \\
+&&\frac{8\pi ^2(a-b)\text{  }\Delta  (\Delta -1) (2 \Delta -1)}{3}\int \text{  }\rho ^2 dx ,
\label{F_matter_general}
\end{eqnarray}
}
where $e(B)$ and $e(W)$ refer to white and black nodes as in \fref{Quiver_Laba_BW}.
Then we impose the constraint $ \sum _{i \in e(B)}  \delta y_{i} +  \sum _{i \in e(W)}  \delta y_{i}=0$
and we compute the Euler-Lagrange equations for $\rho$ and $\delta y$ .
We find
\begin{equation}
\left \{
\begin{array}{lcr}
\delta y_W = 0 & &\\
\delta y_B = \frac{4 a k \pi ^2 x (2-\Delta ) (1-\Delta )}{\mu }&&\quad \quad-\frac{\mu }{2 a k \pi  (2-\Delta )} <x<\frac{\mu }{2 a k \pi  (2-\Delta )}\\
\rho = \frac{\mu }{16 a \pi ^3 \Delta  (2-\Delta ) (1-\Delta )}&&
\end{array}
\right.
\end{equation}
Out of this region we have
\begin{equation}
\left \{
\begin{array}{lcr}
\delta y_W = 0 & &\\
\delta y_B = -2 \pi (1-\Delta)&&\quad \quad  -\frac{\mu }{4 a k \pi  (1-\Delta )}<x<-\frac{\mu }{2 a k \pi  (2-\Delta )}\\
\rho = \frac{4 a k \pi  x (1-\Delta )+\mu }{16 A \pi ^3 (1-\Delta ) \Delta ^2}&&
\end{array}
\right.
\end{equation}
and
\begin{equation}
\left \{
\begin{array}{lcr}
\delta y_W = 0 & &\\
\delta y_B = 2 \pi (1-\Delta)&&\quad \quad \quad \,\,\frac{\mu }{2 a k \pi  (2-\Delta )}<x<  \frac{\mu }{4 a k \pi  (1-\Delta )}\\
\rho = -\frac{4 a k \pi  x (1-\Delta )-\mu }{16 a \pi ^3 (1-\Delta ) \Delta ^2}&&
\end{array}
\right.
\end{equation}
By integrating the piecewise function $\rho$ over the domain where it is non-vanishing we obtain
\begin{equation}
\frac{F^2}{N^3} =
\frac{64 a^2 k \pi ^2 \Delta  (1-\Delta )^2(2-\Delta )^2}{9 (4-3 \Delta )} ,
\end{equation}
which agrees with the result we got from the geometry.

\bigskip

\subsection{Eight extremal points}

\subsubsection{Family 1: $L^{a,b,a}_{(0,\ldots,0,k,-2k||k,0,\ldots,0)}$}

\label{section_8_points_family_1}

We continue our exploration considering a more involved family of geometries with toric diagrams with eight extremal points. The 4-vectors giving the corners of the toric diagram are given in matrix form in \eref{matrix_toric_8_corners_1}. The corresponding toric diagram is given in \fref{matrix_toric_8_corners_1}.

\begin{equation}
\left(
\begin{array}{cccccccc}
\ \ v_1 \ \ & \ \ v_2 \ \ & \ \ v_3 \ \ & \ \ v_4 \ \ & \ \ v_5 \ \ & \ \ v_6 \ \ & \ \ v_7 \ \ & \ \ v_8 \ \ \\
 0 & 1 & 1 & b-1 & b-1 & b & 0 & a \\
 0 & -1 & 1 & -1 & 1 & 0 & 0 & 0 \\
 0 & 0 & 0 & 0 & 0 & 0 & 1 & 1 \\
 1 & 1 & 1 & 1 & 1 & 1 & 1 & 1
\end{array}
\right)
\label{matrix_toric_8_corners_1}
\end{equation}

\begin{figure}[h]
\begin{center}
\includegraphics[width=7.5cm]{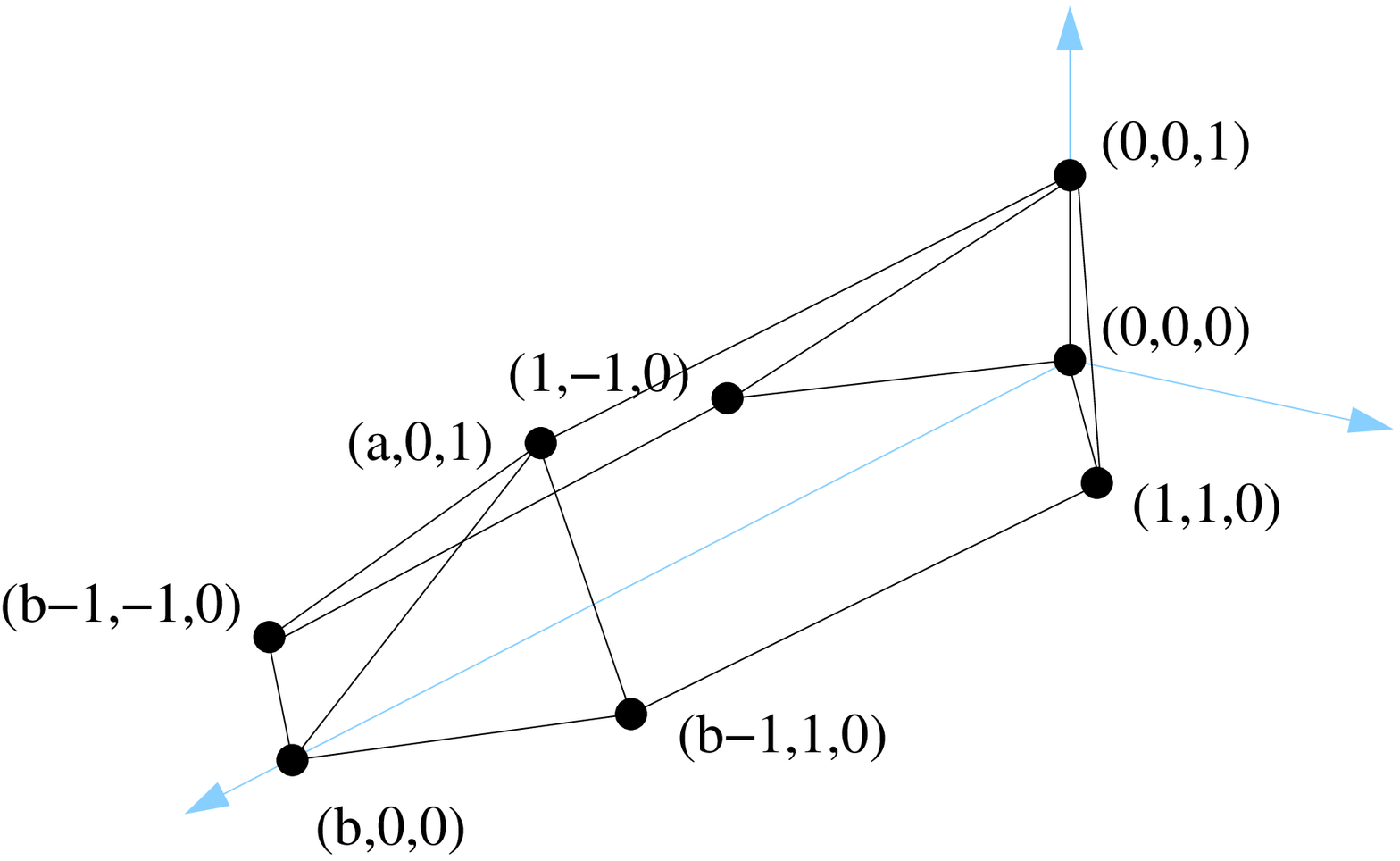}
\caption{Toric diagram for the $L^{aba}_{(0,\ldots,0,k,-2k||k,0,\ldots,0)}$ family with $k=1$.}
\label{toric_8_corners_1}
\end{center}
\end{figure}

This class of models is generated by the lifting algorithm by choosing
the $P(\beta)$ and the $Q(\alpha)$ as
\begin{equation}
\begin{array}{|c|cc|}
\hline 
\ {\rm Type} \ & \ {\rm Multiplicity} \ & \ {\rm Value} \ \\
 \hline
 P & 1 & 0\\
P & b-2 & k\\
P & 1 & 2k\\
Q & a & k\\
\hline
\end{array}
\end{equation}
The resulting CS levels are
\beq
\vec{k}=(0,\ldots,0,k,-2k||k,0,\ldots,0) .
\eeq

\subsection*{Geometric computation}

The $Z_{\rm MSY}$ function in terms of the Reeb vector is
\begin{eqnarray}
&&Z_{\rm MSY}=
\frac{8}{b_3\left(b_1^2-b_2^2\right) \left(b_2^2-\left(b_3-4\right){}^2\right)\left(\left(b_1+b \left(b_3-4\right)-a b_3\right){}^2-b_2^2\right)}
\nonumber \\
&&
(8 b_2^2-b^2 (b_1-b_3+4) (b_3-4)^2+b_1 (b_1-a b_3) ((a+2) b_3-8)+b_3 ((a-2) b_2^2+a^2 (b_3-4) b_3) \nonumber \\
&&
-b (b_3-4) (b_1^2+b_2^2+2 a (b_3-4) b_3-2 b_1 ((a+a) b_3-4))).
\end{eqnarray}
By exploiting the symmetries of toric diagram we can parametrize the components of the Reeb vector as
\begin{equation}
b_1=2 (b(1-\Delta )+a \Delta ),\quad\quad b_2=0,\quad\quad b_3=4 \Delta, 
\end{equation}
where $\Delta$ has a simple relation to the R-charges of fields in the quiver as it will be shown below. The volume function then becomes
\begin{equation}
\text{Vol}(Y_7)=\frac{\pi^4(b+2)(1-\Delta )+a \Delta }{96 k (1-\Delta )^2\text{  }(b(1-\Delta )+a \Delta )^2\Delta } .
\label{Z_8_corners_1}
\end{equation}
Extremizing it, we obtain
{\small
\begin{eqnarray}
\Delta&&=
\frac{1}{12} \left(9+\frac{2 a}{b-a}+\frac{5 a}{b-a+2}\right)+\frac{1}{12 f^{1/3}}\left(9 b (b+2)+2 a \left(a \left(2+\frac{4}{b-a}-\frac{25}{b-a+2}\right)-9\right)\right)\nonumber \\&&+\frac{f^{1/3}}{12 (a(a+2) +(b-2a) (b+2))},
\end{eqnarray}}
where we have defined
{\small\begin{eqnarray}
f&=&
8 a^6-24 a^5 (-2+b)+81 a b^2 (2+b)^3+24 a^4 \left(-32-22 b+b^2\right)-27 a^2 b \left(24+68 b+34 b^2+3 b^3\right)
\nonumber 
\\
&+&a^3 \left(280+1956 b+1074 b^2+19 b^3\right)+3 \left(-72 b^3-108 b^4-54 b^5-9 b^6 +2 \sqrt{3g}\right),
\end{eqnarray}}
and
{\small\begin{eqnarray}
g&=&
a^2 (b-a)^2 (b+2) (b-a+2)^2 
(12 a^3 (b-6) (b+10)-4 a^4 (b+18)-9 b^2 (b+2) (4+(b-28) b)\nonumber \\
&+&2 a (b-14) b (b (11 b+52)-4)-3 a^2 (b (b (2+7 b)-524)+24)).
\end{eqnarray}}
As in the previous family of theories, these models exhibit highly non-trivial values of the R-charges.

\subsection*{Free energy computation}

\begin{figure}[h]
\begin{center}
\includegraphics[width=11cm]{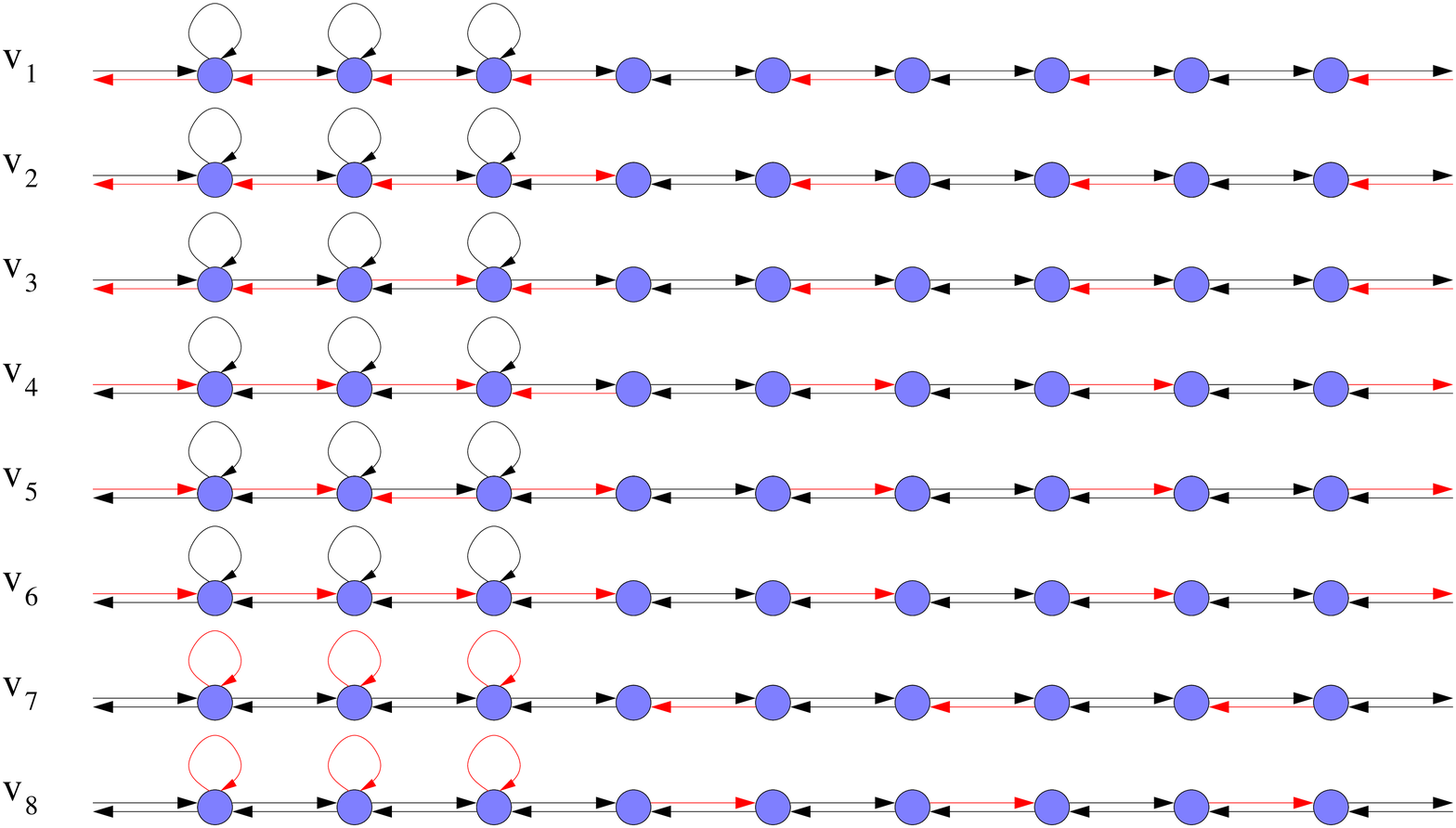}
\caption{Perfect matchings associated to the eight corners of the toric diagram given in 
\eref{toric_8_corners_1}. 
Red arrows indicate the chiral fields associated with edges in the perfect matching.}
\label{first_family_8_points_PM_from_the_quiver}
\end{center}
\end{figure}

We will now recover this complicated structure from the field theory computation of the free energy at large-$N$. The eight extremal perfect matchings are associated to chiral fields as in \fref{first_family_8_points_PM_from_the_quiver}. 

As in previous examples, we use marginality of the superpotential and symmetries to parametrize the R-charges of the extremal perfect matchings in terms of a single parameter $\Delta$ as follows
\begin{equation}
\Delta _{1,6}=\frac{2 (1-\Delta )^2}{((b+2)(1-\Delta )+a \Delta )},\quad \quad
\Delta _{2,3,4,5}=\frac{(1-\Delta ) (b (1-\Delta )+a \Delta )}{2 ((b+2)(1-\Delta )+a \Delta)},\quad \quad\Delta _{7,8}=\Delta .
\end{equation}

The CS contribution to the free energy is
\begin{equation}
\frac{F_{\text{CS}}}{N^{3/2}}= \frac{k}{2 \pi}\int{\rho \left(
\delta y _{b-a-1,b-a} - \delta y _{b-a,b-a+1} 
\right) x dx} .
\end{equation}
The matter contribution is given by the general expression \eref{F_matter_general}. After enforcing the constraint $\sum_{i=1}^{b+1} \delta y_{i,i+2}=0$, we solve the Euler-Lagrange equations to obtain
\begin{equation}
\left \{
\begin{array}{lcrcccl}
\rho= \frac{4 k \pi  x (1-\Delta )+\mu }{16 \pi ^3 (1-\Delta ) \Delta  ((b +2)(1-\Delta )+a\text{  }\Delta )}
& \ \ \ & 
-\frac{\mu }{4 k \pi  (1-\Delta )} & < & x & < & 
-\frac{\mu }{2 k \pi  (b (1-\Delta )+a \Delta )}
\\
\rho =\frac{\mu }{16 \pi ^3 (1-\Delta ) \Delta  (b (1-\Delta )+a \Delta )}
& &
-\frac{\mu }{2 k \pi  (b (1-\Delta )+a \Delta )}
& < & x &
< & \frac{\mu }{2 k \pi  (b (1-\Delta )+a \Delta )}
\\
\rho=- \frac{4 k \pi  x (1-\Delta )-\mu }{16 \pi ^3 (1-\Delta ) \Delta  ((b +2)(1-\Delta )+a\text{  }\Delta )}
& &
\frac{\mu }{2 k \pi  (b (1-\Delta )+a \Delta )}
& < & x & < & 
\frac{\mu }{4 k \pi  (1-\Delta )}
\end{array}
\right.
\end{equation}

The free energy is then
\begin{equation}
{F\over N^{3/2}}=
\frac{64 k \pi ^2 (1-\Delta )^2 \Delta  (b (1-\Delta )+a \Delta )^2}{9 ((b+2)(1-\Delta )+a \Delta )},
\end{equation}
in agreement with the geometric computation \eref{Z_8_corners_1}.

\subsubsection{Family 2: $L^{a,b,a}_{(-k_1,0,\ldots,0,k_{a+b-2X},0,\ldots,0,-k_{a+b-2Y},k_{a+b-2Y+1},0,\ldots,0)}$}

\label{section_8_points_family_2}

\begin{figure}[h]
\begin{center}
\includegraphics[width=7.5cm]{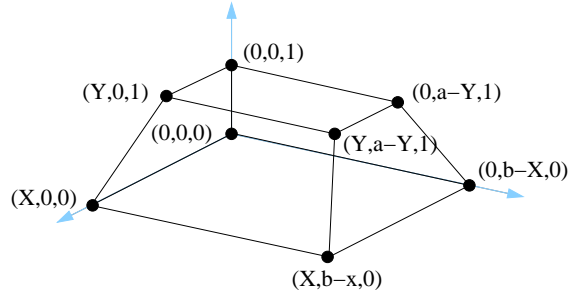}
\caption{Toric diagram for the $L^{a,b,a}_{(-k_1,0,\ldots,0,k_{a+b-2X},0,\ldots,0,-k_{a+b-2Y},k_{a+b-2Y+1},0,\ldots,0)}$ family with $k=1$.}
\label{toric_8_corners_2}
\end{center}
\end{figure}

In this section we study a second family with 8 extremal points.
This family generates the toric diagram in Figure \ref{toric_8_corners_2}. We can arrange the $4d$ vectors generating the diagram in a matrix form
\begin{equation} \label{matrix_toric_8_corners_2}
\left(
\begin{array}{cccccccc}
\ \ v_1 \ \ & \ \ v_2 \ \ & \ \ v_3 \ \ & \ \ v_4 \ \ & \ \ v_5 \ \ & \ \ v_6 \ \ & \ \ v_7 \ \ & \ \ v_8 \ \ \\
 0 & X & X & 0 & 0 & Y & Y & 0 \\
 0 & 0 & b-X & b-X & 0 & 0 & a-Y & a-Y \\
 0 & 0 & 0 & 0 & 1 & 1 & 1 & 1 \\
 1 & 1 & 1 & 1 & 1 & 1 & 1 & 1
\end{array}
\right)
\end{equation}

This geometry follows from $L^{a,b,a}$ by using the lifting algorithm with

\begin{equation}
\begin{array}{|c|cc|}
\hline 
\ {\rm Type} \ & \ {\rm Multiplicity} \ & \ {\rm Value} \ \\
 \hline
P&X&0\\
P &b-X&k\\
Q&Y&0\\
Q&a-Y&k\\
\hline
\end{array}
\end{equation}
The resulting CS levels are

\beq
\vec{k}=(-k_1,0,\ldots,0,k_{a+b-2X},0,\ldots,0,-k_{a+b-2Y},k_{a+b-2Y+1},0,\ldots,0),
\eeq
where the subindices indicate the position of the non-zero entries in the vector $\vec{k}$, which only take values $\pm k$. We will focus on the case in which $b>a$ and $X>Y$. In the notation of \eref{kvec_notation}, we can thus distinguish two possibilities 

\beq
\begin{array}{rccl}
\bullet \ & b>X>a>Y & \ \ \to \ \ & L^{a,b,a}_{(-k_1,0,\ldots,0,k_{a+b-2X},0,\ldots,0||0,\ldots,0,-k_{a+b-2Y},k_{a+b-2Y+1},0,\ldots,0)},
\\ \\
\bullet \ & b>a>X>Y & \to & L^{a,b,a}_{(-k_1,0,\ldots,0||0,\ldots,0,k_{a+b-2X},0,\ldots,0,-k_{a+b-2Y},k_{a+b-2Y+1},0,\dots,0)}.
\end{array}
\eeq
The case $b>a>Y>X$ can be studied in a completely analogous way.

\subsection*{Geometric computation}

For this class of models, $Z_{\rm MSY}$ takes the form

\begin{equation}
Z_{\rm MSY}=
\frac{4 \left(X \left(4-b_3\right)+Y b_3\right) \left((b-X) \left(b_3-4\right)-a b_3+Y b_3\right)}{b_1 b_2 \left(b_3-4\right) b_3 \left(b_1+X \left(b_3-4\right)-Y b_3\right) \left(b_2+(b-X) \left(b_3-4\right)-a b_3+Y b_3\right)}.
\end{equation}
Imposing marginality of the superpotential and symmetries, we have
\begin{equation}
\begin{array}{ccl}
b_1 & = & \frac{1}{2} (4 X-4 X \Delta +4 Y \Delta ) \\
b_2 & = & \frac{1}{2} (4 b-4 X+4 a \Delta -4 b \Delta +4 X \Delta -4 Y \Delta ) \\
b_3 & = & 4 \Delta
\end{array}
\end{equation}
and the volume becomes
\begin{equation}
\text{Vol}(Y_7) =
\frac{\pi^4}{48 k (1-\Delta ) \Delta  ((b-X) (1-\Delta )+(a-Y) \Delta ) (X (1-\Delta )+Y \Delta )} .
\end{equation}
As in previous examples, it is straightforward to find the value of $\Delta$ that minimizes the volume analytically. The resulting expression is not terribly illuminating, so we do not quote it here.

\subsection*{Free energy computation}

\begin{figure}[h]
\begin{center}
\includegraphics[width=11cm]{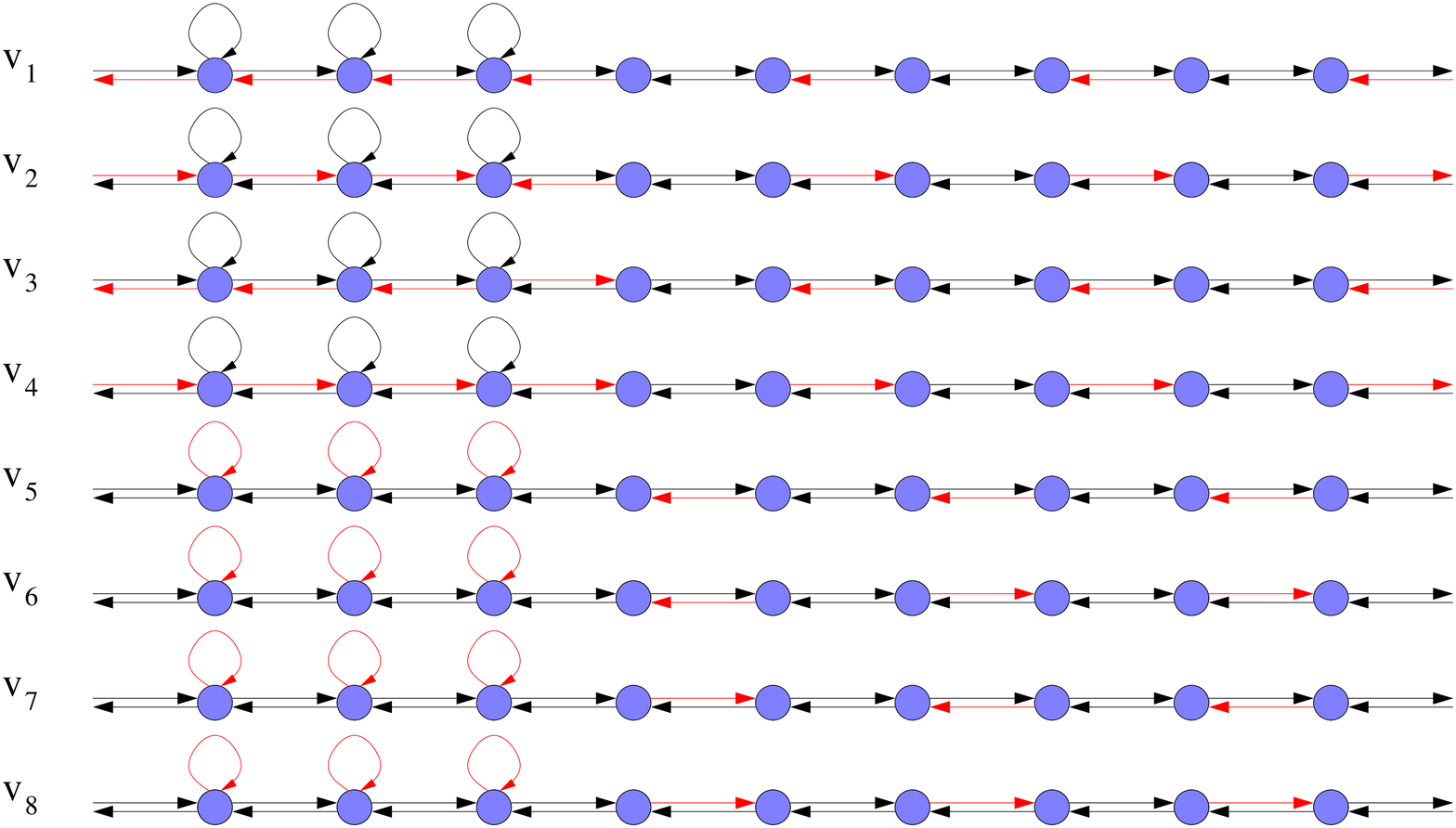}
\caption{Perfect matchings associated to the eight corners of the toric diagram given in 
\eref{toric_8_corners_2}, with $X=Y=1$. 
Red arrows indicate the chiral fields associated with edges in the perfect matching.}
\label{PM_Laba_8_points_2}
\end{center}
\end{figure}

The eight extremal perfect matchings are associated to chiral fields as in 
\fref{PM_Laba_8_points_2}.\footnote{The specific values $X=Y=1$ used in \fref{PM_Laba_8_points_2} have been chosen for illustration purposes only. In this case, the two CS contributions $k_{a+b-2X}$ and $k_{a+b-2Y}$ correspond to the same entry in $\vec{k}$ and cancel each other, reducing the theories to $L^{a,b,a}_{(-k,0,\ldots,0||0,\ldots,0,k_{a+b-1},0)}$. Determining the perfect matchings for the $X>Y$ regime considered in this section is straightforward.} The free energy for the gauge theory can be written by distinguishing two different $\delta y$'s
as
\begin{equation}
F_{\text{CS}} = \frac{k}{2 \pi} \int \rho x \left( (b-X) \delta y _{1} + (a-x) \delta y_{2} \right) dx,
\end{equation}
and 
{\small
\begin{eqnarray}
F_{\text{matter}} &=& 
(\!b\!-\!X\!) F_{\text{bif}}(\!1\!-\!\Delta , \delta y_1)\!+\!
(\!a\!-\!Y\!)F_{\text{bif}}(\Delta, \delta y_2)\!+\!
(\!X\!-\!1\!) F_{\text{bif}}(\!1\!-\!\Delta, \delta y_3)\!
+\!Y F_{\text{bif}} (\Delta,\delta y_4)\nonumber \\
&+&
F_{\text{bif}}( 1-\Delta, (b-X) \delta y_1+  (a-Y) \delta y_2+  (X-1) \delta y_3+  Y \delta y_4)
+
(b-a) F_{\text{adj}}(2 \Delta),\nonumber 
\\
\end{eqnarray}}
where $F_{\text{bif}}$ and $F_{\text{adj}}$ are the contributions to the free energy of 
a couple of bifundamental anti-bifundamental and of an adjoint field.
By computing the saddle point equations we find
\begin{equation}
\frac{F}{N^{3/2}}=
\frac{32}{9} k \pi ^2 (1-\Delta ) \Delta  ((b-X) (1-\Delta )+(a-Y) \Delta ) (X (1-\Delta )+Y \Delta ),
\end{equation}
which matches the volume computation.

\section{Free Energy as a Quartic Function in R-charges}

\label{section_quartic}

In this section we would like to discuss the existence of a geometrical formula 
capable of reproducing the free energy in terms of the charges of the perfect matchings similar to the one derived in \cite{Butti:2005vn} for SCFT$_4$'s.

\bigskip

\subsection{4d preliminaries}

Before continuing our study of the free energy of SCFT$_3$'s, it is useful discuss related questions in 4d. The number of degrees of freedom of an $\mathcal{N}=1$ SCFT in 4d is counted by the central charge $a$, which can be determined in terms of superconformal R-charges \cite{Anselmi:1997am,Intriligator:2003jj} as follows

\beq
a={3\over 32}(3\Tr R^3 - \Tr R).
\label{a_field_theory}
\eeq
Furthermore, for SCFTs on D3-branes, $\Tr R=0$ and \eref{a_field_theory} becomes purely cubic. In \cite{Martelli:2005tp,Butti:2005vn,Lee:2006ru,Eager:2010yu}, 
it has been shown that $a$ also admits a similar cubic expression based on the underlying geometry, which takes the form
\begin{equation}
a_\text{geom} = \frac{9}{32} \sum_{i,j,k} |\langle v_i,v_j,v_k \rangle| R_i R_j R_k,
\label{a_geometric}
\end{equation}
where the $v_i$ are the 3-dimensional vectors defining the extremal points of the two dimensional toric diagram and the $R_i$ are the $R$-charges of the perfect matchings associated to $v_i$.

While \eref{a_geometric} is written in terms of quantities that allow a direct contact with geometry, it is important to keep in mind that perfect matchings are indeed identified with GLSM fields which, in turn, can be found in purely field theoretic terms starting from the gauge theory and computing its moduli space. Equation \eref{a_geometric} can also be obtained by rewriting the inverse of the volume of the 5d Sasaki-Einstein base of the corresponding toric CY$_3$, which takes the form

\beq
{\rm Vol} (Y_5)= \sum_{i}\frac{
\langle v_{i-1},v_i,v_{i+1}\rangle }{\langle b,v_{i-1},v_i\rangle
\langle b,v_i,v_{i+1}\rangle },
\label{V_toric_Y5}
\eeq
where the $v_i$ vectors are the 3-vectors with the coordinates of extremal points in the toric diagram and $b=(b_1,b_2,3)$ is the Reeb vector. Due to the Calabi-Yau condition, we can take $v_i=(\tilde{v}_i,1)$, with $\tilde{v}_i$ a 2-vector. $\langle \cdot ,\cdot ,\cdot \rangle$ is the determinant of the resulting $3\times 3$ matrix.

\bigskip

\subsection{Free energy in 3d}

In Section \ref{section_free_energy}, we have explained how to compute the free energy of SCFT$_3$'s. Furthermore, we have shown that its value agrees with the geometric computation in various infinite classes of theories. It is natural to wonder whether a simple expression for the free energy, similar in spirit to \eref{a_field_theory} exist in 3d. The main obstacle for going into this direction is the absence of anomalies associated to continuous symmetries in 3d. Having said this, the similarity between the volume formulas \eref{MSvol}, \eref{Vol_toric_Y7} and \eref{V_toric_Y5} suggest that an expression in terms of R-charges of perfect matchings, i.e. of GLSM fields, analogous to \eref{a_geometric} might exist. The most naive generalization of \eref{a_geometric} to 3d takes the form

\begin{equation} 
F_\text{geom}^2 =\frac{1}{6} \sum_{i,j,k,l}|\langle v_i,v_j,v_k,v_l \rangle|  \Delta_i \Delta_j \Delta_k \Delta_l ,
\label{F_geometric}
\end{equation}
where we have used $\Delta_i$ instead of $R_i$ to match the notation we have been using for SCFT$_3$'s. Remarkably, it has been observed in \cite{Amariti:2011uw} that this formula reproduces the free energy of several theories. Even in specific models for which \eref{F_geometric} does not give the correct result, it has been possible to introduce additional terms such that the free energy is still given by a quartic formula in the R-charges of extremal perfect matchings. Interestingly, in all the theories considered in \cite{Amariti:2011uw} the corrections to \eref{F_geometric} seem to be connected to the existence of {\it internal lines} in the toric diagram, i.e. lines connecting extremal points that do not live on edges or faces. 

\bigskip

\subsection{Quartic formulas for $L^{a,b,a}_{\vec{k}}$ theories}

We now go over all the classes of models considered in Section \ref{section_infinite_families} and show that, in all of them, the free energy can be written as a quartic function of the R-charges of extremal perfect matchings. It is important to emphasize that this agreement holds off-shell, i.e. even before extremizing the free energy.

For the first two families of $L^{a,b,a}_{(0,\dots,0||k,-k,\dots,k,-k)}$ and $L^{a,b,a}_{(k,0,\dots,0||-k,0,\dots,0)}$ theories, discussed in Sections \ref{section_4_points_family_1} and 
\ref{section_6_points_family_1}, the free energy is exactly reproduced by \eref{F_geometric}. Geometrically, these two families distinguish themselves from the others in that their toric diagrams do not contain internal lines, i.e. all lines connecting corners of the toric diagram live on edges or external faces.

\begin{figure}[h]
\begin{center}
\includegraphics[width=15cm]{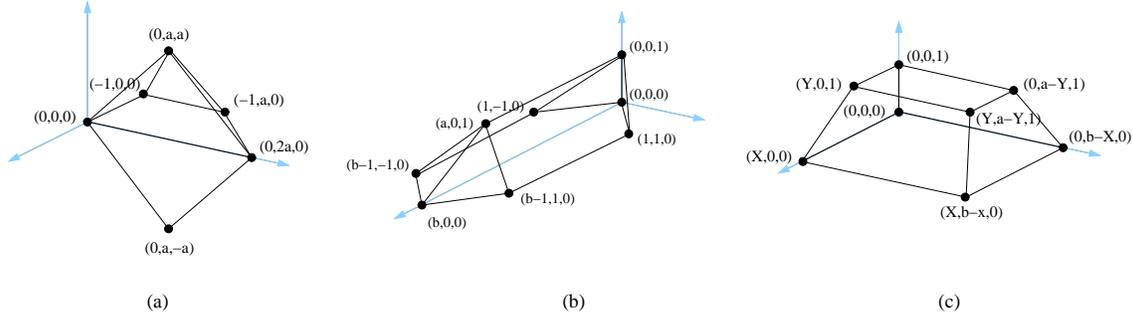}
\caption{Toric diagrams for: a) $L^{a,2a,a}_{(0,\dots,0,-2k||k,k,-k,k,-k,\dots,k,-k,k)}$, b) $L^{a,b,a}_{(0,\ldots,0,k,-2k||k,0,\ldots,0)}$ and c) $L^{a,b,a}_{(-k_1,0,\ldots,0,k_{a+b-2X},0,\ldots,0,-k_{a+b-2Y},k_{a+b-2Y+1},0,\ldots,0)}$.}
\label{toric_internal_lines}
\end{center}
\end{figure}

The remaining families require corrections to \eref{F_geometric}, but can still be recast in quartic form. We reproduce the toric diagrams in \fref{toric_internal_lines} for quick reference. Contrary to the first two families of geometries, these models contain internal lines connecting extremal perfect matchings in the toric diagram. The corrections are given by 

{\small
\beq
\begin{array}{lccccl}
\bullet \ L^{a,2a,a}_{(0,\dots,0,-2k||k,k,-k,k,-k,\dots,k,-k,k)} & : & \ & \Delta F^2 & = & -2 a^2 (\Delta_1^2 \Delta_6^2+\Delta_4^2\Delta_5^2)+ 4 a^2  \Delta_1 \Delta_6 \Delta_4 \Delta_5 \\ \\
\bullet \ L^{a,b,a}_{(0,\ldots,0,k,-2k||k,0,\ldots,0)}& : & & \Delta F^2 & = & -2 a (\Delta_1^2 \Delta_8^2+\Delta_6^2\Delta_7^2)+ a \Delta_1 \Delta_6 \Delta_7 \Delta_8 \\ \\
\bullet \ L^{a,b,a}_{(-k_1,0,\ldots,k_{a+b-2X},0,\ldots,-k_{a+b-2Y},k_{a+b-2Y+1},0,\ldots)} & : & & \Delta F^2 & = & -4 X(a-Y)
\left(\Delta _3^2 \Delta _5^2+\Delta _4^2 \Delta _6^2+\Delta _1^2 \Delta _7^2 \right. \\
& & & & & \left. + \Delta _2^2 \Delta _8^2- 4\Delta _2 \Delta _4 \Delta _6 \Delta _8-4\Delta _1 \Delta _3 \Delta _5 \Delta _7 \right)
\end{array}
\eeq
}
For the last family we have restricted to the case $aX=bY$ because it exhibits additional symmetries that simplify the computation.

All these models contain terms of the form $\Delta_i^2 \Delta_j^2$. Their coefficients seem to admit some simple expression in terms of the toric diagram. For example, for the $L^{a,2a,a}_{(0,\dots,0,-2k||k,k,-k,k,-k,\dots,k,-k,k)}$ models we have

\begin{eqnarray}
\Delta_1^2 \Delta_6^2 & \rightarrow & 
-2 a^2 = -4 \frac{
|\langle v_2,v_3,v_1,v_6\rangle |
|\langle v_3,v_5 ,v_1,v_6\rangle |
|\langle v_5,v_2 ,v_1,v_6\rangle |
}
{
|\langle v_2,v_3,v_5,v_1\rangle |
|\langle v_2,v_3,v_5,v_6\rangle |
},
\nonumber\\
\Delta_1^2 \Delta_6^2 & \rightarrow & 
-2 a^2 = -4 \frac{
|\langle v_2,v_3,v_4,v_5\rangle |
|\langle v_3,v_6 ,v_4,v_5\rangle |
|\langle v_6,v_2 ,v_4,v_5\rangle |
}
{
|\langle v_2,v_3,v_6,v_4\rangle |
|\langle v_2,v_3,v_6,v_5\rangle |
}.
\end{eqnarray}
Identical expressions, even including the same $(-4)$ numerical factor, apply for the $\Delta_i^2 \Delta_j^2$ terms for the $L^{a,b,a}_{(0,\ldots,0,k,-2k||k,0,\ldots,0)}$ family. 

\bigskip

\subsection{Towards a general quartic formula}

\label{section_towards_quartic}

The previous examples lead us to some conjectures regarding the possible structure of a general quartic formula. It appears that there are two possible types of corrections to \eref{F_geometric}, which arise in the presence of internal lines in the toric diagram:

\medskip

\begin{itemize}
\item[{\bf 1)}] A correction proportional to $\Delta_i^2 \Delta_j^2$, whenever the line connecting extremal points $i$ and $j$ of the toric diagram is internal.
\item[{\bf 2)}] A correction proportional to $\Delta_i\Delta_j \Delta_k\Delta_l$, whenever the lines connecting the extremal points $i$ and $j$, and $k$ and $l$ are both internal.
\end{itemize}

\medskip

\begin{figure}[h]
\begin{center}
\includegraphics[width=6cm]{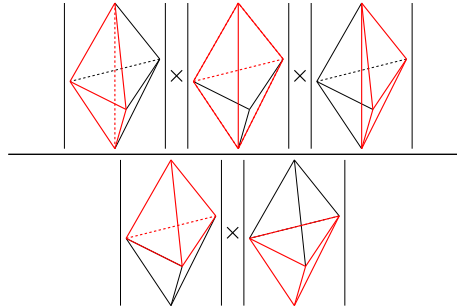}
\caption{Graphical representation for the numerical coefficient of the $\Delta_i^2 \Delta_j^2$ term in the free energy as given by 
\eref{Delta_i_Delta_j_coefficient}.
}
\label{M111like}
\end{center}
\end{figure}

Furthermore, based on the examples, it is possible to conjecture an explicit expression for the numerical coefficient multiplying the corrections of type {\bf (1)}. If a line connecting extremal point intersects an internal triangle, then one takes the product of the volumes of the three possible tetrahedra ($V_1$, $V_2$ and $V_3$) whose vertices are the two endpoints of the line and a pair of vertices of the triangle, and divide it by the product of the volumes of two tetrahedra ($V_4$ and $V_5$) given by the triangle and each of the endpoints of the line. The corresponding correction to the free energy is of the form
\beq
\Delta F_g^2 = -4 \frac{V_1 V_2 V_3}{V_4 V_5} \Delta_i ^2 \Delta_j^2  .
\label{Delta_i_Delta_j_coefficient}
\eeq
The $(-4)$ prefactor is universal whenever an internal line intersects a triangle. This prescription admits a nice graphical representation as shown in \fref{M111like}

More complicated situations can be obtained by triangulation. For example, if a line line ($i \rightarrow j$)  crosses a polygon formed by four extremal points of the toric diagram ($\{v_1,v_2,v_3,v_4\}$) we generate four tetrahedra with vertices being $i$ and $j$ and the other two on the polygon, whose volumes are: $V_1=|\langle v_1, v_2,v_i,v_j\rangle |$,  $V_2=|\langle v_2, v_3,v_i,v_j\rangle |$, $V_3=|\langle v_3 ,v_4,v_i,v_j\rangle |$ and $V_4=|\langle v_4 ,v_1,v_i,v_j\rangle |$. On the other hand, we construct two other volumes made out of tetrahedra with a single vertex being $i$ or $j$: $V_5=|\langle v_1 v_2,v_3,v_i\rangle |+|\langle v_2 v_3,v_4,v_i\rangle |$ and $V_6=|\langle v_1 v_2,v_3,v_j\rangle |+|\langle v_2 v_3,v_4,v_j\rangle |$. The resulting correction is 
\begin{equation}\label{triangulation}
\Delta F_g^2  = - \frac{ V_1 V_2 V_3+V_2 V_3 V_4+V_3 V_4 V_1+V_4 V_1 V_2}{V_5 V_6} \Delta_i^2 \Delta_j^2 .
\end{equation}

In Appendix \ref{appendix_non_Laba}, we present additional examples beyond the $L^{a,b,a}_{\vec{k}}$ theories, providing further support for the general ideas advocated in this section. Eventually, we expect a geometric identity that systematically re-expresses the volume function \eref{MSvol} as a quartic function in the volumes $\text{Vol}(\Sigma_i)$ of elements in the basis of 5-cycles.

\section{Toric Duality and the Free Energy}

\label{section_toric_duality}

A corollary of the lifting algorithm of Section \ref{section_lifting_algorithm} is that individual permutations of the $P_\beta$ and $Q_\alpha$ subsets of 5-branes lead to gauge theories with different CS couplings but the same CY$_4$ manifold as their mesonic moduli space. This invariance suggests that the corresponding gauge theories are dual. Theories that share the same toric moduli space, in any dimensions, go under the general denomination of {\it toric duals} \cite{Feng:2000mi}. In fact, some of these permutations, those that exchange a pair of adjacent $P_\beta$ and $Q_\alpha$ branes, indeed correspond to the 3d version of Seiberg duality discussed in \cite{Franco:2008um,Amariti:2009rb,Franco:2009sp,Davey:2009sr,Giveon:2008zn,Niarchos:2008jb}.
It was then shown in \cite{Herzog:2010hf,Amariti:2011uw,Gulotta:2011vp,Gulotta:2012yd} that the large-$N$ free energy is preserved under this duality.

We now show that, as expected for dual theories, the free energy is invariant under permutations of $P_\beta$ and $Q_\alpha$ branes. Let us start from the CS contribution which, following \eref{free_energy_contributions}, is proportional to
\begin{equation} 
\sum k_i y_i = \sum(p_{i-1}-p_i) y_i = \sum p_i \delta y_i .
\end{equation}

Following the separation of the $p_i$ into two sets $\{p_i\} = \{P_\beta,Q_\alpha\}$ we also divide the $\delta y_i$ as $\{\delta y_i\} = \{\delta y_\beta,\delta y_\alpha\}$. The CS contribution then becomes
\begin{equation} \label{Fcs}
F_{\text{CS}} = \frac{N^{3/2}}{2 \pi}\int \rho \, x 
\left(
\sum_{\{\alpha\}} Q_\alpha \delta y_\alpha +
\sum_{\{\beta\}} P_\beta \delta y_\beta 
\right)
dx .
\end{equation}
Defining\footnote{Here $2 \Delta=\Delta^{(+)}$ in  \ref{free_energy_contributions} while $\Delta^{(-)}$ has been set to zero. We assume that R-charges can be parametrized as in \fref{quiver_Laba}. This is the case in all the infinite families of models considered in this paper and can occur whenever symmetries of the toric diagram impose further constraints on the R-charges of extremal perfect matchings, which are then translated into constraints on the R-charges of quiver fields.}
\begin{eqnarray}
F_{\text{bif}}(\Delta, \delta y)  & =&  -(1-\Delta ) \int  \rho^2 \left( \delta y^2-\frac{4}{3} \pi ^2\Delta (2-\Delta )\right) dx \nonumber \\
F_{\text{adj}}(\Delta) & = & \frac{2}{3} \pi ^2 (1-\Delta ) (2-\Delta ) \Delta \int  \rho ^2 dx
\end{eqnarray}
the matter contribution  is given by
\begin{equation} \label{fmat}
F_{\text{matter}} = 
\sum_{\{\alpha\}} F_{\text{bif}} \left(1-\Delta, \delta y_\alpha\right)
+
\sum_{\{\beta\}} F_{\text{bif}} \left(\Delta, \delta y_\beta \right)+
(b-a) F_{\text{adj}}(2 \Delta) .
\end{equation}

Next, we consider the action of two arbitrary elements $S_a$ and $S_b$ of the symmetric group acting on $Q_\alpha$ and $P_\beta$, respectively. We see that both (\ref{Fcs}) and (\ref{fmat}) are preserved if we simultaneously act with the same permutation actions $S_a$ and $S_b$ on $\delta y_\alpha$'s and $\delta y_\beta$'s. This shows that, as expected from the invariance of the moduli space, the large-$N$ free energy is preserved.

\bigskip

\section{Conclusions}

\label{section_conclusions}

Remarkable progress in understanding SCFT$_3$'s on M2-branes and in the field theoretic calculation of the number of degrees of freedom in these SCFTs has taken place in recent years. One of the main goals of this paper has been to accumulate a large body of evidence, in the form of infinite classes of theories, explicitly showing the expected agreement \cite{Martelli:2011qj,Jafferis:2011zi} between the volume of the Sasaki-Einstein horizon of the probed CY$_4$ cone and the free energy of the dual field theory computed on a round $S^3$. The infinite families of models we investigated in Section \ref{section_infinite_families} belong to the $L^{a,b,a}_{\vec{k}}$ class, and their corresponding gauge theories have generically $\mathcal{N}=2$ SUSY and the same vector-like quivers and superpotentials  of D3-branes on real cones over $L^{a,b,a}$ manifolds. These theories also include CS couplings, encoded in the vector $\vec{k}$, which dictate how the parent CY$_3$ manifold is lifted to a CY$_4$. Our results provide non-trivial checks of the AdS$_4$/CFT$_3$ correspondence for infinite families of gauge theories and it is a step towards a general proof of the equivalence between the $Z_{\rm {MSY}}$-minimization and the $F$-maximization.

The infinite families we studied were generated with the aid of a lifting algorithm we introduced in Section \ref{section_lifting_algorithm}, which is based on the Type IIB realization of these theories and allows us to efficiently generate the CY$_4$ geometries for $L^{a,b,a}_{\vec{k}}$ theories.

Our results are similar to the equivalence between $Z_{\rm {MSY}}$-minimization and $a$-maximization in 4d \cite{Martelli:2005tp}-\cite{Eager:2010yu}, whose proof for toric theories relies on the existence of a geometric formula for the central charge, $a_\text{geom}$, that is cubic in the R-charges of extremal perfect matchings \cite{Martelli:2005tp,Butti:2005vn}. This follows crucially from the relation between the geometry of extremal perfect matchings and triangle anomalies in field theory \cite{Benvenuti:2006xg}. Despite the absence of anomalies in 3d, the similarity of the geometric expression for the horizon volumes between the 3d and 4d case makes it natural to expect that a geometric expression  for the free energy $F_\text{geom}^2$, quartic in the R-charges of extremal perfect matchings, exists in 3d. In Section \ref{section_quartic}, we have shown that this expression exists for all the infinite families of theories we studied. Furthermore, the correspondence is valid even before extremization. Counting with an infinite catalogue of examples has allowed us to make various conjectures regarding the general form of the quartic formula. These ideas were tested in additional, non-$L^{a,b,a}_{\vec{k}}$ models in Appendix \ref{appendix_non_Laba}, verifying that they indeed agree with the volume. We find all these results are encouraging and make us expect that it is possible to rewrite the volume formula as a quartic expression in volumes of 5-cycles. It would be very interesting to show that such a formula exists and to give a systematic prescription for writing it based on the toric data. 

In the future, it is certainly desirable to prove the equivalence between $Z_{\rm {MSY}}$-minimization and $F$-maximization for general toric geometries. A more modest objective is to prove the equivalence within some sub-class of theories, such as the $L^{aba}_{\vec{k}}$ models. Our results go a long way in this direction, but we had to be specific about the choice of CS levels in order to perform the calculations. It would be interesting to find an efficient procedure for dealing with a generic choice of CS levels. First, one should manage to find the volume of $Y_7$ for an arbitrary choice gauge theory data, $a$, $b$ and $\vec{k}$. Hilbert Series techniques \cite{Davey:2009sr,Davey:2009qx} seem to be a promising direction for achieving this goal. On the field theory, one should compute the free energy, i.e. solve the corresponding Euler-Lagrange equations, for a generic distribution of CS levels. Computing the free energy from a Fermi gas, as proposed in \cite{Marino:2011eh} for $\mathcal{N}\geq 3$ theories,  is perhaps a more promising approach, since no matrix model techniques are needed. 

We conclude with some comments on toric duality. In Section \ref{section_toric_duality}, we have shown that toric duals generated by permuting 5-branes in the type IIB realization of $L^{a,b,a}_{\vec{k}}$ theories preserve the large-$N$ free energy without fractional branes, i.e. for all the ranks of the gauge group being equal. It is natural to expect that at finite $N$ the precise ranks of the gauge groups might become important for the duality. This issue can be investigated by rewriting the free energy as in \cite{Benini:2011mf}, using the formalism of \cite{vandeBult}.  Imposing the correct balancing conditions on the integrals associated to the free energy of the candidate dual phases, it should be possible to determine the number of fractional branes required by duality.

\section*{Acknowledgements}

We would like to thank K. Intriligator, C. Klare and M. Siani for useful discussions. A. A. is supported by UCSD
grant DOE-FG03-97ER40546. The work of S. F. was supported by the US DOE under contract number DE-AC02-76SF00515 and by the U.K. Science and Technology Facilities Council (STFC).


\bigskip
\bigskip

\appendix

\section{Non-$L^{a,b,a}_{\vec{k}}$ Theories and Quartic Formulas}

\label{appendix_non_Laba}

In this appendix, we provide additional evidence supporting our proposals of Section \ref{section_towards_quartic}. To do so, we consider two families of theories that do not fit within the $L^{a,b,a}_{\vec{k}}$ classification. In the first class of geometries, the toric diagram is given by

\begin{equation}
\left(
\begin{array}{ccccc}
\ \ v_1 \ \ & \ \ v_2 \ \ & \ \ v_3 \ \ & \ \ v_4 \ \ & \ \ v_5 \ \ \\
 1 & -1 & 0 & 0 & 0 \\
 0 & -1 & 1 & 0 & 0 \\
 0 & 0 & 0 & X_1 & -X_2 \\
 1 & 1 & 1 & 1 & 1
\end{array}
\right)
\end{equation}
where $X_i>0$. These theories have already been studied in \cite{Amariti:2011uw}, where it has been shown that the geometrical free energy is given by \eref{F_geometric} plus the following correction
\begin{eqnarray} \label{correM111}
\Delta F_\text{geom}^2 &= & - 4
\frac{\left(X_1+X_2\right)^3}{9 X_1 X_2} \Delta_4^2 \Delta_5^2.
\end{eqnarray}
This correction is associated to an internal line connecting $v_4$ and $v_5$ in the toric diagram. The numerical coefficient of this correction is in perfect agreement with our proposal \eref{Delta_i_Delta_j_coefficient}. Using it, we obtain
\begin{equation} \label{correM1112}
\Delta_4^2 \Delta_5^2 \rightarrow
-4 \frac{ 
|\langle v_1 v_2 v_4 v_5 \rangle | |\langle  v_2 v_3 v_4 v_5 \rangle | |\langle  v_3 v_1 v_4 v_5\rangle |
}
{|\langle  v_1 v_2 v_3 v_4  \rangle | |\langle  v_1 v_2 v_3 v_5 \rangle |
}=-4 
\frac{ \left(X_1+X_2\right)^3}{9 X_1 X_2}.
\end{equation}

The final set of models we would like to consider has a toric diagram given by
\begin{equation}
\left(
\begin{array}{cccccccc}
\ \ v_1 \ \ & \ \ v_2 \ \ & \ \ v_3 \ \ & \ \ v_4 \ \ & \ \ v_5 \ \ & \ \ v_6 \ \ \\
 X_1 & -X_2 & 0 & 0 & 0 & 0 \\
 0 & 0 & Y_1 & -Y_2 & 0 & 0 \\
 0 & 0 & 0 & 0 & Z_1 & -Z_2 \\
 1 & 1 & 1 & 1 & 1 & 1
\end{array}
\right)
\end{equation}
with $X_i$, $Y_i$, $Z_i>0$. In this case, a quartic expression for the free energy also exists, and it is given by \eref{F_geometric} plus the rather non-trivial correction

{\scriptsize
\begin{eqnarray}\label{Q222like}
\Delta F_\text{geom}^2 &=& 
-\frac{2 \left(X_1+X_2\right)^3 Y_1 Y_2 Z_1 Z_2 }{X_1 X_2 \left(Y_1+Y_2\right) \left(Z_1+Z_2\right)}\Delta _1^2 \Delta _2^2
-\frac{2 X_1 X_2 \left(Y_1+Y_2\right)^3 Z_1 Z_2 }{\left(X_1+X_2\right) Y_1 Y_2 \left(Z_1+Z_2\right)}\Delta _3^2 \Delta _4^2
-\frac{2 X_1 X_2 Y_1 Y_2 \left(Z_1+Z_2\right)^3}{\left(X_1+X_2\right) \left(Y_1+Y_2\right) Z_1 Z_2} \Delta _5^2 \Delta _6^2
\nonumber  
\\
&+&\frac{4 \left(X_1+X_2\right) \left(Y_1+Y_2\right) Z_1 Z_2}{Z_1+Z_2} \Delta _1 \Delta _2 \Delta _3 \Delta _4
+\frac{4 \left(X_1+X_2\right) Y_1 Y_2 \left(Z_1+Z_2\right)}{Y_1+Y_2} \Delta _1 \Delta _2 \Delta _5 \Delta _6
\nonumber  
\\
&+&\frac{4 X_1 X_2 \left(Y_1+Y_2\right) \left(Z_1+Z_2\right)}{X_1+X_2} \Delta _3 \Delta _4 \Delta _5 \Delta _6 .
\end{eqnarray}}

It is possible to check that the $\Delta_1^2 \Delta_2^2$, $\Delta_3^2 \Delta_4^2$ and $\Delta_5^2 \Delta_6^2$ terms are indeed in agreement with \eref{triangulation}, including its $(-1)$ prefactor.


\end{document}